\documentstyle[twoside,psfig,epsfig]{article}

\catcode`\@=11
\long\def\@makefntext#1{
\protect\noindent \hbox to 3.2pt {\hskip-.9pt
$^{{\eightrm\@thefnmark}}$\hfil}#1\hfill}               

\def\@makefnmark{\hbox to 0pt{$^{\@thefnmark}$\hss}}    

\def\ps@myheadings{\let\@mkboth\@gobbletwo
\def\@oddhead{\hbox{}
\rightmark\hfil\eightrm\thepage}
\def\@oddfoot{}\def\@evenhead{\eightrm\thepage\hfil
\leftmark\hbox{}}\def\@evenfoot{}
\def\sectionmark##1{}\def\subsectionmark##1{}}

\oddsidemargin=\evensidemargin
\newcounter{sectionc}\newcounter{subsectionc}\newcounter{subsubsectionc}
\renewcommand{\section}[1] {\vspace{12pt}\addtocounter{sectionc}{1}
\setcounter{subsectionc}{0}\setcounter{subsubsectionc}{0}\noindent
        {\tenbf\thesectionc. #1}\par\vspace{5pt}}
\renewcommand{\subsection}[1] {\vspace{12pt}\addtocounter{subsectionc}{1}
        \setcounter{subsubsectionc}{0}\noindent
        {\bf\thesectionc.\thesubsectionc. {\kern1pt \bfit #1}}\par\vspace{5pt}}
\renewcommand{\subsubsection}[1] {\vspace{12pt}\addtocounter{subsubsectionc}{1}
        \noindent{\tenrm\thesectionc.\thesubsectionc.\thesubsubsectionc. 
        {\kern1pt \tenit #1}}\par\vspace{5pt}}

\newcounter{appendixc}
\newcounter{subappendixc}[appendixc]
\newcounter{subsubappendixc}[subappendixc]
\renewcommand{\thesubappendixc}{\Alph{appendixc}.\arabic{subappendixc}}
\renewcommand{\thesubsubappendixc}
        {\Alph{appendixc}.\arabic{subappendixc}.\arabic{subsubappendixc}}

\renewcommand{\appendix}[1] {\vspace{12pt}
        \refstepcounter{appendixc}
        \setcounter{figure}{0}
        \setcounter{table}{0}
        \setcounter{lemma}{0}
        \setcounter{theorem}{0}
        \setcounter{corollary}{0}
        \setcounter{definition}{0}   
        \setcounter{equation}{0}
        \renewcommand{\thefigure}{\Alph{appendixc}.\arabic{figure}}
        \renewcommand{\thetable}{\Alph{appendixc}.\arabic{table}}
        \renewcommand{\theappendixc}{\Alph{appendixc}}
        \renewcommand{\thelemma}{\Alph{appendixc}.\arabic{lemma}}
        \renewcommand{\thetheorem}{\Alph{appendixc}.\arabic{theorem}}
         \renewcommand{\thedefinition}{\Alph{appendixc}.\arabic{definition}}
        \renewcommand{\thecorollary}{\Alph{appendixc}.\arabic{corollary}}
        \renewcommand{\theequation}{\Alph{appendixc}.\arabic{equation}}
        \noindent{\tenbf Appendix \theappendixc #1}\par\vspace{5pt}}
\newcommand{\subappendix}[1] {\vspace{12pt}
        \refstepcounter{subappendixc}
        \noindent{\bf Appendix \thesubappendixc. {\kern1pt \bfit #1}}
        \par\vspace{5pt}}
\newcommand{\subsubappendix}[1] {\vspace{12pt}
        \refstepcounter{subsubappendixc}
        \noindent{\rm Appendix \thesubsubappendixc. {\kern1pt \tenit #1}}
        \par\vspace{5pt}}

\newcommand{\textlineskip}{\baselineskip=13pt}
\newcommand{\smalllineskip}{\baselineskip=10pt}

\def\eightcirc{
\begin{picture}(0,0)
\put(4.4,1.8){\circle{6.5}}
\end{picture}}
\def\eightcopyright{\eightcirc\kern2.7pt\hbox{\eightrm c}}

\renewenvironment{thebibliography}[1]
        {\frenchspacing
         \ninerm\baselineskip=11pt
         \begin{list}{\arabic{enumi}.}
        {\usecounter{enumi}\setlength{\parsep}{0pt}
         \setlength{\leftmargin 12.7pt}{\rightmargin 0pt} 
         \setlength{\itemsep}{0pt} \settowidth
        {\labelwidth}{#1.}\sloppy}}{\end{list}}

\newcounter{itemlistc}
\newcounter{romanlistc}
\newcounter{alphlistc}
\newcounter{arabiclistc}

\newcommand{\fcaption}[1]{
        \refstepcounter{figure}
        \setbox\@tempboxa =
\hbox{\footnotesize Fig.~\thefigure. #1}
        \ifdim \wd\@tempboxa > 5in
           {\begin{center}
           \parbox{5in}{\footnotesize\smalllineskip Fig.~\thefigure. #1}
            \end{center}}
        \else
             {\begin{center}
             {\footnotesize
Fig.~\thefigure. #1}
              \end{center}}
        \fi}

\newcommand{\tcaption}[1]{
        \refstepcounter{table}
        \setbox\@tempboxa = \hbox{\footnotesize Table~\thetable. #1}
        \ifdim \wd\@tempboxa > 5in
           {\begin{center}
        \parbox{5in}{\footnotesize\smalllineskip Table~\thetable. #1}
            \end{center}}
        \else
             {\begin{center}
             {\footnotesize Table~\thetable. #1}
              \end{center}}
        \fi}

\def\@citex[#1]#2{\if@filesw\immediate\write\@auxout
        {\string\citation{#2}}\fi
\def\@citea{}\@cite{\@for\@citeb:=#2\do
        {\@citea\def\@citea{,}\@ifundefined
        {b@\@citeb}{{\bf ?}\@warning
        {Citation `\@citeb' on page \thepage \space undefined}}
        {\csname b@\@citeb\endcsname}}}{#1}}

\newif\if@cghi
\def\cite{\@cghitrue\@ifnextchar [{\@tempswatrue
        \@citex}{\@tempswafalse\@citex[]}}
\def\citelow{\@cghifalse\@ifnextchar [{\@tempswatrue
        \@citex}{\@tempswafalse\@citex[]}}
\def\@cite#1#2{{$\null^{#1}$\if@tempswa\typeout
        {IJCGA warning: optional citation argument
        ignored: `#2'} \fi}}

\def\pmb#1{\setbox0=\hbox{#1}
        \kern-.025em\copy0\kern-\wd0
        \kern.05em\copy0\kern-\wd0
        \kern-.025em\raise.0433em\box0}

\def\fnt#1#2{\footnotetext{\kern-.3em
        {$^{\mbox{\scriptsize #1}}$}{#2}}}
        
\def\fpage#1{\begingroup
\voffset=.3in
\thispagestyle{empty}\begin{table}[b]\centerline{\footnotesize #1}
        \end{table}\endgroup}
              
\font\tenrm=cmr10
\font\tenit=cmti10
\font\tenbf=cmbx10
\font\bfit=cmbxti10 at 10pt
\font\ninerm=cmr9

\font\eightrm=cmr8

\def\be{\begin{equation}}
\def\ee{\end{equation}}

\def\qed{\hbox{${\vcenter{\vbox{                        
   \hrule height 0.4pt\hbox{\vrule width 0.4pt height 6pt
   \kern5pt\vrule width 0.4pt}\hrule height 0.4pt}}}$}}

\begin{document}
\setlength{\textheight}{7.7truein}  


\normalsize\textlineskip
\thispagestyle{empty}
\setcounter{page}{1}



\fpage{1}
\centerline{\bf THE NEUTRALINO MASS: CORRELATION WITH THE CHARGINOS}
\vspace*{0.37truein}
\centerline{\footnotesize M\"{U}GE BOZ}
\baselineskip=12pt
\centerline{\footnotesize\it Physics  Department, Hacettepe University}
\baselineskip=10pt
\centerline{\footnotesize\it Ankara, 06532, Turkey }
\vspace*{10pt}

\centerline{\footnotesize NAMIK  K. PAK}
\baselineskip=12pt
\centerline{\footnotesize\it Physics Department, Middle East Tecnical  University}
\baselineskip=10pt
\centerline{\footnotesize\it Ankara, 06531, Turkey }
\vspace*{10pt}

\begin{abstract}
\noindent
As the fundamental SU(2) supersymmetric parameters can be 
determined in the chargino sector, and the 
remaining fundamental parameters of the minimal supersymmetric 
extensions of the standard model  can 
be analyzed in the neutralino sector, the two sectors 
can be correlated via these parameters.
We have shown that for the CP conserving case, 
the masses of all the neutralinos can be determined in terms of 
the chargino masses,
and $\tan\beta$. In this case the neutralino masses are quite insensitive to the
variations of $\tan\beta$; they change by about  $\%15$ when $\tan\beta$
varies in the range from 5 to 50.
In the CP violating case, the neutralino masses are found to be 
quite sensitive to the variations of the CP violating phase.
For the heavier neutralinos the dependence of the masses to the CP violating phase
show complementary behaviour at CP violating points.\\
\end{abstract}
\section{Introduction and Summary}
The Lagrangian of the Minimal Supersymmetric  Standard Model 
(MSSM) contains various mass parameters which are not necassarily real~\cite{Dugan}.
The phases of these parameters appear in several   CP violating processes 
such as the electric dipole moments~\cite{Ibrahim}, 
the decays and mixings of mesons~\cite{DemirOlive},
the Higgs phenomenology~\cite{Pilaftsis1,Boz2,BozPak},
and the chargino/neutralino systems~\cite{Choi1,Choi2,Choi3}.

One of the simplest sectors in supersymmetric (SUSY) theories 
is that of  the charginos. The $2\times 2$ chargino mass matrix 
\begin{eqnarray}
M_{\chi}=\left(\begin{array}{c c}  M_{2} & \sqrt{2} M_{W} \sin \beta  \\
\sqrt{2} M_{W} \cos  \beta & \mu  \end{array}\right)~,
\end{eqnarray}
is built up by the SU(2) gaugino,
and the Higgsino mass parameters, $M_2$ and $\mu$, respectively, and the ratio
$\tan\beta=v_2/v_1$ of the expectation values of the two neutral
Higgs fields which break the electroweak symmetry.

In the CP violating theories  $M_2$, and 
$\mu$ can be 
complex. However, by the reparametrization of  the fields $M_2$ can be taken as 
real and positive, so that the remaining  non-trivial phase can
be attributed to the  $\mu$ parameter.
We define,
\begin{eqnarray}
\mu=|\mu|e^{i \varphi_\mu} 
\end{eqnarray}
The chargino mass matrix $M_{\chi}$  can be diagonalized  by 
the following transformation:
\begin{eqnarray}
\label{def}
{\cal{U}}^{*} M_{\chi} {{\cal{V}}^{-1}} = \mbox{Diag} 
(M_{ {\chi}^+_{1}}, M_{ {\chi}^+_{2}})~,
\end{eqnarray}
with the chargino mass  eigenvalues $m^2_{\tilde{\chi}^+_{1,2}}$: 
\begin{eqnarray}
m^2_{\tilde{\chi}^+_{1,2}}
   =\frac{1}{2}\left[M^2_2+|\mu|^2+2 M^2_W \mp \Delta_\chi \right]
\end{eqnarray}
where  
\begin{eqnarray}
\Delta_\chi  &=&\bigg[(M^2_2-|\mu|^2-2 M_W^2 \cos 2 \beta)^2\nonumber\\ 
               &+& 
               8 M_W^2 ( M_2 ^2 \cos^2 \beta + |\mu|^2 \sin^2 \beta 
               + M_2 |\mu| \sin 2 \beta \cos\varphi_\mu) \bigg]^{1/2}
\end{eqnarray}
gives the difference between the two chargino masses 
($M_{\chi^+_2}^2- M_{\chi^+_1}^2$).

For given $\tan\beta$, the fundamental SUSY parameters $M_2$ and $|\mu|$
can be derived from these two masses~\cite{Choi2,Moultaka98}.
The sum and the difference of the chargino masses lead to the following equations 
involving  $M_2$ and $|\mu|$:
\begin{eqnarray}
\label{M2}
M_2^2+ |\mu|^2 =  M^2_{\chi^+_1} + M^2_{\chi^+_2}-2 M^2_W~, 
\end{eqnarray} 
\begin{eqnarray}
\label{eq7}
   M_2^2|\mu|^2 - 2 M^2_W \sin 2 \beta \cos \varphi_{\mu} M_2 |\mu|  +
   (M^4_W \sin^{2} 2 \beta - M^2_{\chi^+_1} M^2_{\chi^+_2})=0~.
\end{eqnarray}
The solution of (\ref{eq7}) is given as:
\begin{eqnarray}
\label{m2mu}
M_2 |\mu| = M^2_W \cos \varphi_{\mu} \sin 2\beta \pm \sqrt {M_{\chi^+_1}^2 M_{\chi^+_2}^2-
M_W^4 \sin^2  2 \beta \sin^2 \varphi_{\mu}}~.
\end{eqnarray}
From (6) and (8) one obtains the following solutions for $M_2$ and $|\mu|$:
\begin{eqnarray}
\label{M2mu}
2 M_2^2  &=&   (M^2_{\chi^+_1}+M^2_{\chi^+_2}- 2 M_W^2) \mp 
\sqrt{ (M^2_{\chi^+_1}+M^2_{\chi^+_2}- 2 M_W^2)^2 -Q_{\pm}}~,
\end{eqnarray}
\begin{eqnarray}
2 |\mu|^2  &=&   (M^2_{\chi^+_1}+M^2_{\chi^+_2}- 2 M_W^2) \pm 
\sqrt {(M^2_{\chi^+_1}+ M^2_{\chi^+_2}- 2 M_W^2)^2 - Q_{\pm}}~,
\end{eqnarray} 
with
\begin{eqnarray}
Q_{\pm} &=&4 \bigg[M^2_{\chi^+_1} M^2_{\chi^+_2}
+M_W^4 \cos 2 \, \varphi_{\mu} \sin^ 2 2 \beta \nonumber\\ 
&\pm &  2 M_W^2 \cos \varphi_{\mu} \sin 2 \beta 
\sqrt{M^2_{\chi^+_1} M^2_{\chi^+_2} -M_W^4 \sin^2 \varphi_{\mu} \sin^2  2 \beta}\bigg]~, 
\end{eqnarray} 
where the upper signs correspond to $M_2 < |\mu|$ regime, and the lower ones to 
$M_2 > |\mu|$.

Therefore, for given $\tan\beta$,  $M_2$
and $|\mu|$ can be determined in terms of the masses of the charginos 
(${M_{\chi^+_1}}$ and ${M_{\chi^+_2}}$)   by  
using (9), and (10) from which one gets four solutions corresponding to 
different physical scenarios.
For $|\mu|< M_2$,  the lightest chargino has a stronger higgsino-like component
and therefore is referred as higgsino-like~\cite{Choi3,Moultaka98}. 
The solution   $|\mu|>M_2$, corresponding to gaugino-like situation, can be 
readily obtained  by 
the substitutions:
$M_2 \rightarrow |\mu|$, and $\mu \rightarrow \mbox{sign} 
(\mu) \, M_2$~\cite{Choi3,Moultaka99}.

Let us now consider the mass matrix of the neutralino system:
\begin{eqnarray} M_{{\chi}^{0}} \ =\ \left(
\begin{array}{cccc}  M_{1} & 0 & -M_Z s_W  \cos\beta &  M_Z s_W  \sin\beta\\
0 &  M_{2} &  M_Z c_W \cos \beta & -M_Z c_W \sin \beta\\ 
-M_Z s_W \cos \beta &  M_Z c_W \cos \beta & 0 & -\mu \\ 
M_Z s_W \sin \beta & -M_Z c_W \sin \beta &
-\mu & 0 \end{array}\right),
\end{eqnarray}
The main difference of the mass spectra of the  neutralino and chargino
system is the appearence of the SU(2) gaugino mass $M_1$, in the former.

The neutralino mass matrix  can be diagonalized as  follows:
\begin{eqnarray}
{\cal{N}}^{T} \,  M_{{\chi}^{0}} \,  {\cal{N}} =\mbox{Diag} \left(M_{\chi^{0}_4}, \cdots,
M_{\chi^{0}_1}\right)~,
\label{neutmat}
\end{eqnarray}
with ordering 
$M_{\chi^{0}_4} > M_{\chi^{0}_3} > M_{\chi^{0}_2} > M_{\chi^{0}_1}$.

Assuming the  two chargino masses are known, it is possible to express the 
neutralino masses in terms of  these, for given $\tan\beta$. 
In this work, we have obtained the neutralino masses numerically, by using
(\ref{neutmat}). In doing this,  we use   (9) and (10), for given $\tan\beta$.

Complete analytical solutions 
can be derived for the 
neutralino mass eigenvalues 
(\ref{neutmat}) as functions of the SUSY parameters 
for both CP conserving~\cite{GunionHaber},
and  CP violating theories~\cite{Choi3}.
Admitedly, the diagonalization 
of the neutralino mass  matrix is no easy job and the analytic expressions of the 
resulting eigenvalues  are rather lengthy and 
complicated.

However, there are  theoretically well motivated 
assumptions, like for instance the universality of the soft mass parameters,
which could be easily implementable to the system. 
Typically,
the gaugino mass parameter universality at the grand unification (GUT) scale,
leads to the approximate relation~\cite{Moultaka98}:
\begin{eqnarray}
M_1(M_Z) = 5/3 \tan^2 \theta_W  M_2 (M_Z)~.
\end{eqnarray}

Furthermore, there are also very reasonable approximations
to these mass eigenvalues in limiting cases which are sufficiently compact to 
allow a good understanding 
of the analytic dependencies.
For instance, a particularly interesting 
limit is approached when the 
the supersymmetry mass parameters and their
splittings are much larger than 
the electroweak scale $M_{SUSY}^2 >> M_Z^2$.
In this limit 
the neutralino mass eigenvalues can be written in compact 
(approximate) form as~\cite{Choi3}:
\begin{eqnarray}
 M_{\chi^{0}_1}& =& |M_1| + {\cal Z}_1 \bigg[|M_1|+|\mu|\sin 2\beta
         \cos \varphi_\mu \bigg]~,
\end{eqnarray}
\begin{eqnarray}
M_{\chi^{0}_2} & 
=& |M_2| + {\cal Z}_2 \bigg[|M_2|+|\mu|\sin 2 \beta
         \cos \varphi_\mu \bigg]~,
\end{eqnarray}
\begin{eqnarray}
M_{\chi^{0}_3}  &=&|\mu| \bigg [1 - \frac{(1-\sin 2 \beta)}{2} \, ( {\cal Z}_1+{\cal Z} _2)\bigg]\nonumber\\[1mm]
        &+& \frac{(1-\sin 2 \beta)}{2}\, \bigg[ {\cal Z}_1|M_1|
        + {\cal Z}_2 |M_2| \bigg]\cos\varphi_\mu~, 
\end{eqnarray}
\begin{eqnarray}
M_{\chi^{0}_4} & =&|\mu|\bigg 
[1 - \frac{(1+\sin 2 \beta)}{2} \, ( {\cal Z}_1+{\cal Z} _2)\bigg]\nonumber\\[1mm]
        &-& \frac{(1+\sin 2 \beta)}{2} \, \bigg[ {\cal Z}_1 |M_1|
        + {\cal Z}_2 |M_2| \bigg]\cos\varphi_\mu~, 
\end{eqnarray}
where 
\begin{eqnarray}
 {\cal Z}_ 1 =\frac{m^2_Z\, s^2_W}{|M_1|^2-|\mu|^2} \qquad {\rm and} \qquad
 {\cal Z }_2 =\frac{m^2_Z\, c^2_W}{|M_2|^2-|\mu|^2}~.
\end{eqnarray}

The masses of the charginos and neutralinos 
are  interesting  observables  which provide 
clues about the  SUSY-breaking structure of  the 
system~\cite{Tesla}. Therefore, particle masses, SUSY parameters,
the relations between the masses
themselves, the relations between the basic gaugino parameters and  the physical masses,
are important for calculations.
Previous works on the subject include the analysis 
at the lowest order processes~\cite{Moultaka99}, in the on-shell scheme~\cite{Hollik},
and aim to reconstruct  the basic parameters based on  
chargino production~\cite{Choi1,Choi2}.

In this work, our aim is to   obtain the neutralino masses, from those of the charginos
and investigate the effects of the CP violating phase on the masses of the neutralinos,
taking the two chargino masses and $\tan\beta$  as input parameters.

In CP violating theories, the gaugino mass $M_2$, and the Higgsino-Dirac
mass parameter $\mu$ can be complex. However, 
the gaugino mass $M_2$ can be taken to be  real,  
and  hence the phase of  the $\mu$ parameter 
becomes the only non-trivial CP violating phase
in the theory. In this work, we  choose $M_2$ to be real.
which   means $M_1$ to be   real also,
due to  the interrelation between them. Thus, the only 
non-trivial CP violating phase can
be attributed to the $\mu$ parameter.

In the following, we first briefly consider  the CP conserving case,
where we calculate the neutralino masses numerically, 
and analyze their dependence of $\tan\beta$.
Then we turn to the case for which there is CP violation in the theory,
and study  the sensitivity of the neutralino masses to the CP violating phase.

\section{Numerical Analysis}

\subsection{CP conserving case}

In the first part of the analysis,
we set $\varphi_{\mu}=0$ and  
we take the two chargino masses, 
and  $\tan\beta$  as input parameters,
and calculate the neutralino masses $M_{\chi^{0}_i}$.

In our analysis, we fix the heavy chargino mass  as $M_{\chi^+_2}=320~\mbox{GeV}$, 
and choose two different values for the light chargino  mass ($M_{\chi^+_1}$), as 
$\tan\beta$ varies  from 5 to 50.
\begin{figure}[htb]
\vspace*{-2.5truein}
\hspace*{0.3truein}
\centerline{\psfig{file=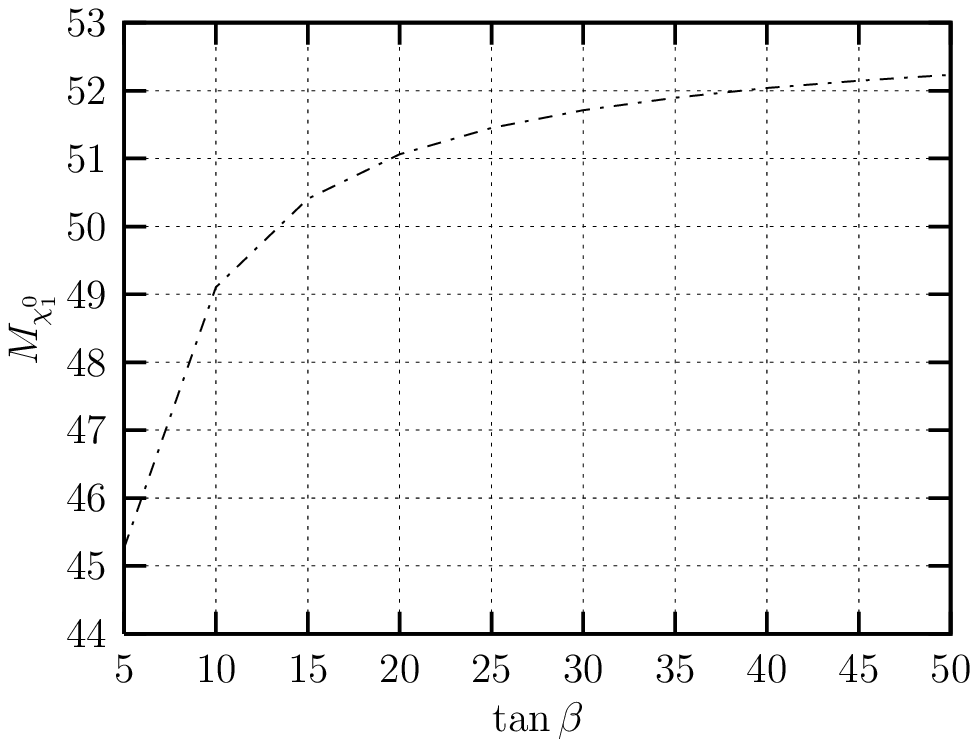,height=7.0in,width=5.0in }
\hspace*{-2.6truein}
\psfig{file=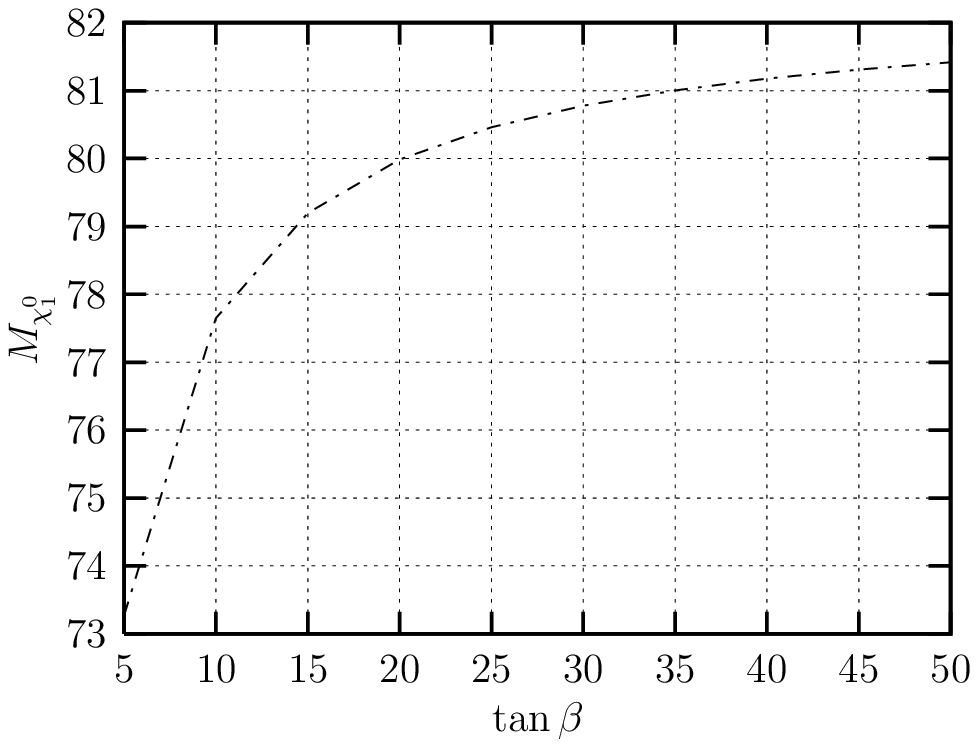,height=7.0in,width=5.0in }}
\vspace*{-2.9truein}
\fcaption{The $\tan\beta$ dependence of 
$M_{{\chi}^{0}_{1}}$, when   
$M_{\chi^+_1}=105~\mbox{GeV}$~(left panel), and   $M_{\chi^+_1}=160~\mbox{GeV}$~(right panel)
for $M_2<|\mu|$.} 
\label{fig1}
\end{figure}
\begin{figure}[htb]
\vspace*{-2.5truein}
\hspace*{0.3truein}
\centerline{\psfig{file=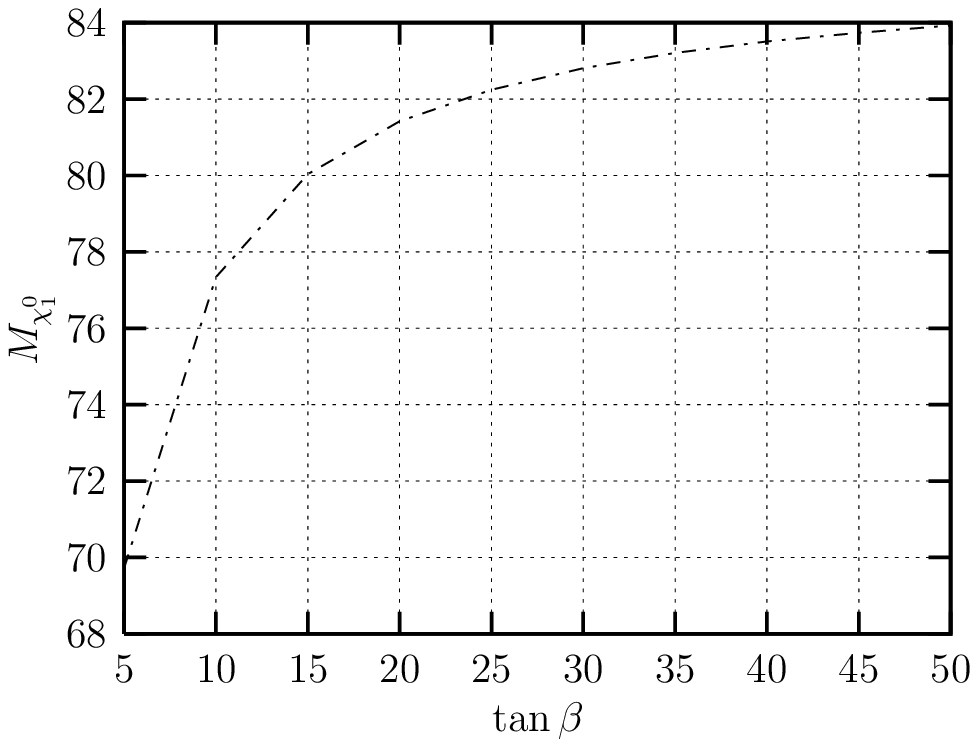,height=7.0in,width=5.0in }
\hspace*{-2.6truein}
\psfig{file=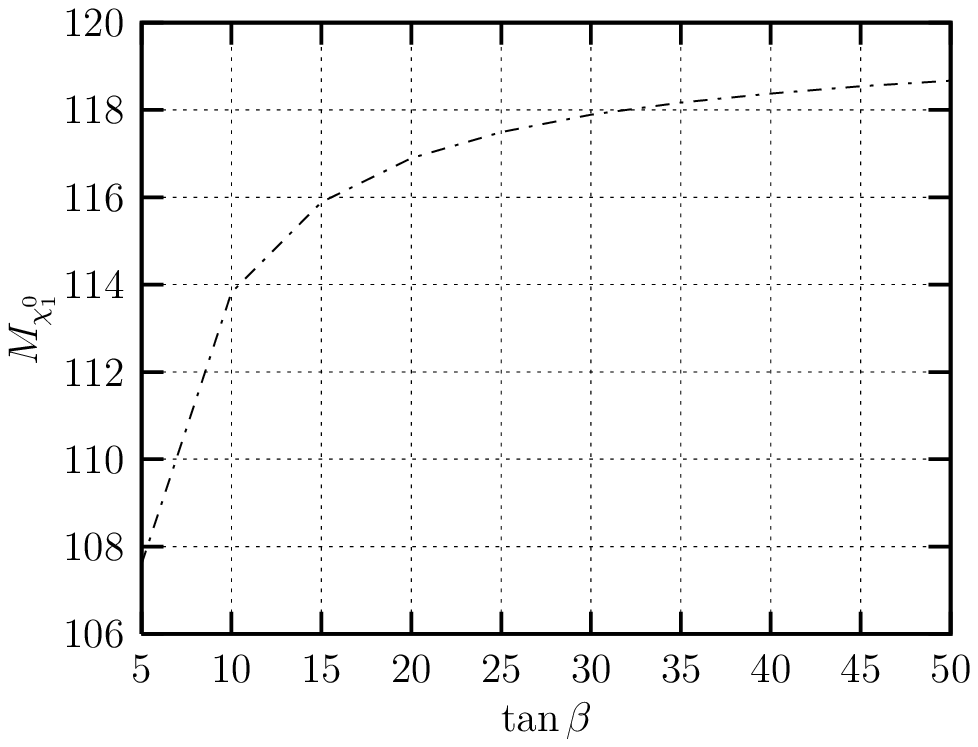,height=7.0in,width=5.0in }}
\vspace*{-2.9truein}
\fcaption{The $\tan\beta$ dependence of 
$M_{{\chi}^{0}_{1}}$, when   
$M_{\chi^+_1}=105~\mbox{GeV}$~(left panel), and   $M_{\chi^+_1}=160~\mbox{GeV}$~(right panel)
for $M_2>|\mu|$.} 
\label{fig2}
\end{figure}

In Figure 1 and Figure 2, we plot the variation of the lightest neutralino mass $M_{{\chi}^{0}_{1}}$
with respect to $\tan\beta$, 
for $M_2 < |\mu|$   and   for $M_2 > |\mu|$ regimes,  respectively.
In both Figures the left panels are for $M_{\chi^+_1}=105~\mbox{GeV}$, and  
the right panels are for $M_{\chi^+_1}=160~\mbox{GeV}$.

It can be seen from both Figures  that  
$M_{{\chi}^{0}_{1}}$ increases with $\tan\beta$, 
at both values of the lightest chargino mass 
($M_{\chi^+_1}=105~\mbox{GeV}$ and   $M_{\chi^+_1}=160~\mbox{GeV}$)
for both $M_2 < |\mu|$   and  $M_2 > |\mu|$ regimes.

One can deduce that when $M_{\chi^+_1}=105~\mbox{GeV}$,
the gaugino and Higgsino Dirac mass  
lie  in the  $M_2 \, (|\mu|) \sim 104-113~\mbox{GeV}$  
and $|\mu|\, (M_2) \sim 299-296~\mbox{GeV}$ intervals, respectively, for $M_2 < |\mu|$ 
($M_2 > |\mu|$). 
When   $M_{\chi^+_1}=160~\mbox{GeV}$, one can again deduce that
$M_2 \, (|\mu|)$ ranges from    164 to 175$~\mbox{GeV}$, whereas   
$|\mu| \, (M_2)$  changes from 297 to 290  $\mbox{GeV}$.

A comparative analysis of Figure 1 and Figure 2 suggest that 
the lightest neutralino mass ($M_{{\chi}^{0}_{1}}$) changes by at most  $\%15$
as  $\tan\beta$ varies from 5 to
 50, thus depicting a low sensitivity.
For instance, the maximal and minimal values of 
$M_{{\chi}^{0}_{1}}$ can be read as 45~$\mbox{GeV}$, and 52~$\mbox{GeV}$, 
at $\tan\beta=5$, and  $\tan\beta=50$, respectively, at $M_{\chi^+_1}=105~\mbox{GeV}$, 
for $M_2 < |\mu|$. 
Similar observations  can be made for the   $M_2 > |\mu|$ regime.
\begin{figure}[htb]
\vspace*{-2.5truein}
\hspace*{0.3truein}
\centerline{\psfig{file=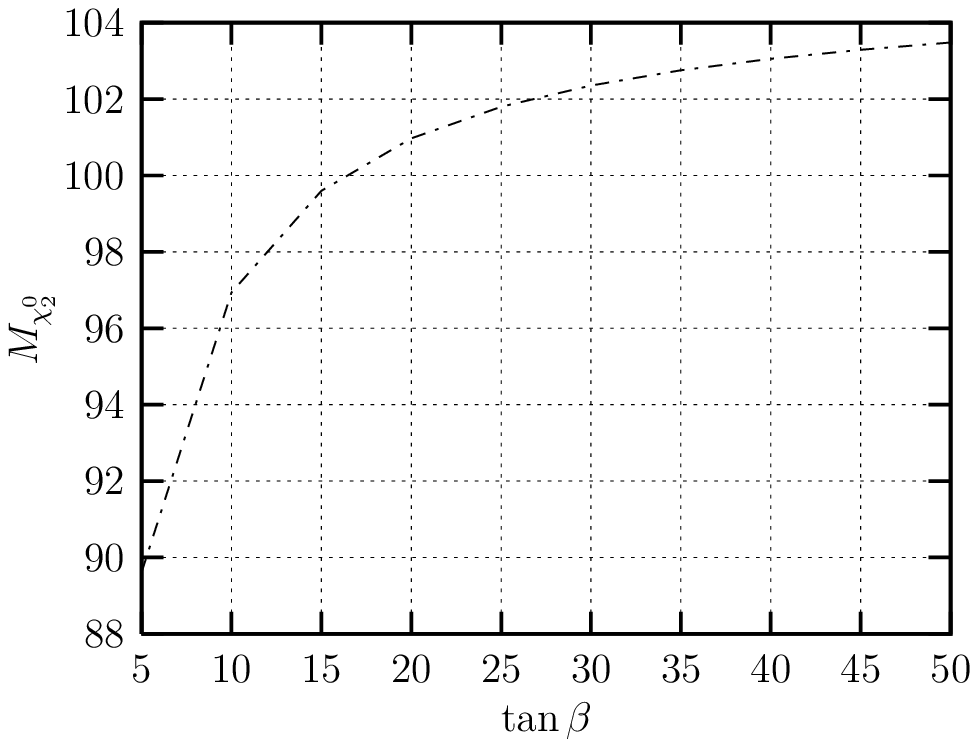,height=7.0in,width=5.0in }
\hspace*{-2.6truein}
\psfig{file=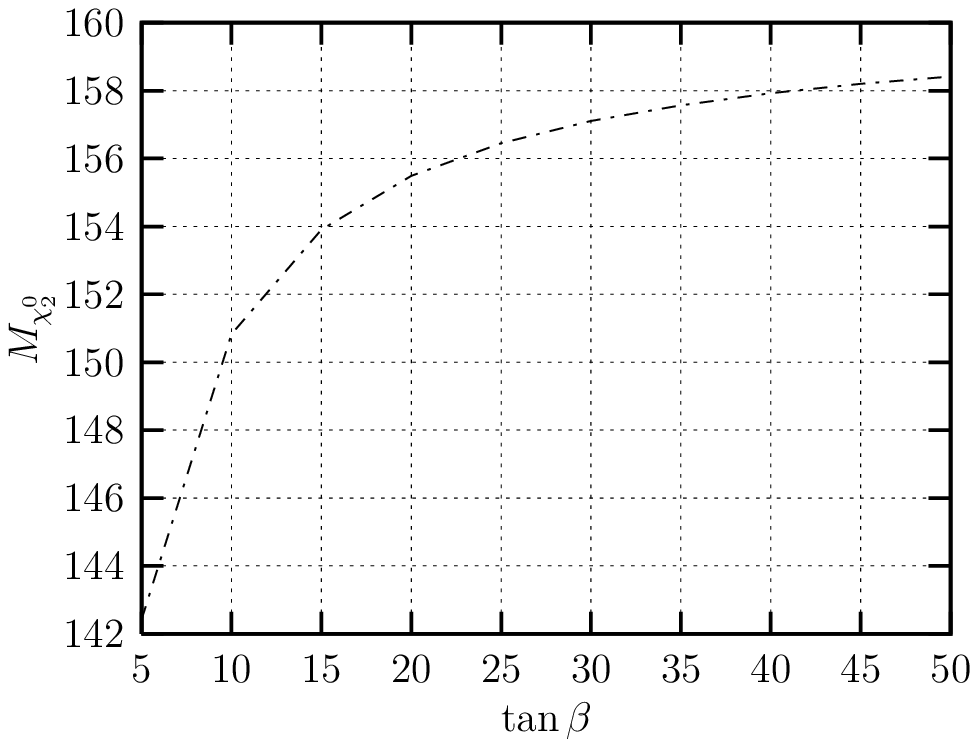,height=7.0in,width=5.0in }}
\vspace*{-2.9truein}
\fcaption{ The $\tan\beta$ dependence of 
$M_{{\chi}^{0}_{2}}$, when   
$M_{\chi^+_1}=105~\mbox{GeV}$~(left panel), and  
$M_{\chi^+_1}=160~\mbox{GeV}$~(right panel) $M_2<|\mu|$.} 
\label{fig3}
\end{figure}
\begin{figure}[htb]
\vspace*{-2.5truein}
\hspace*{0.3truein}
\centerline{\psfig{file=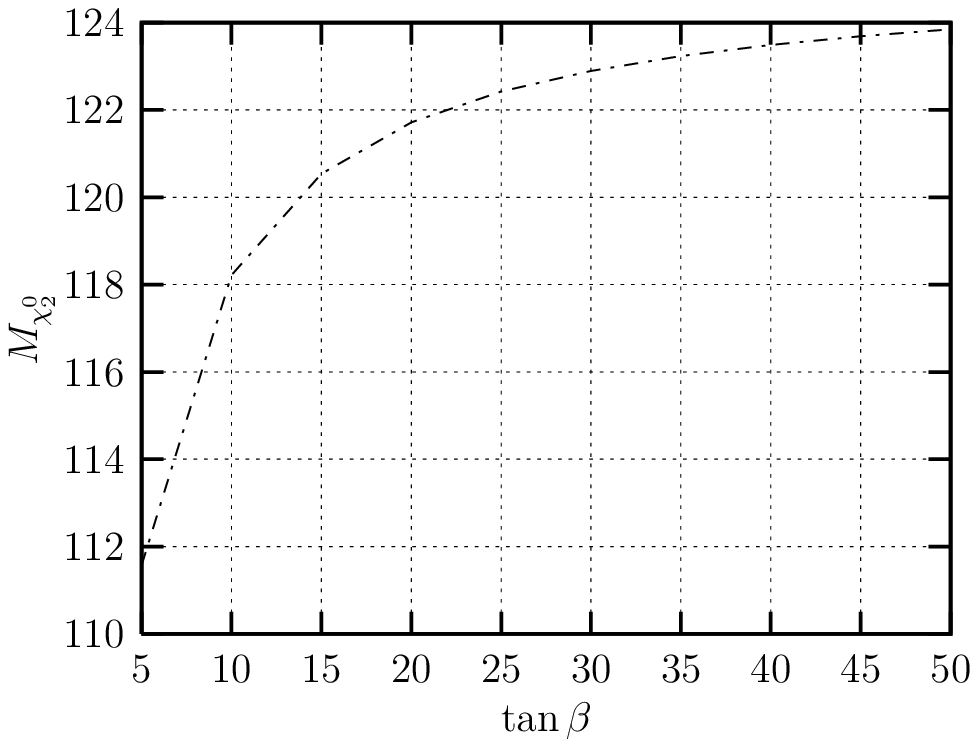,height=7.0in,width=5.0in }
\hspace*{-2.6truein}
\psfig{file=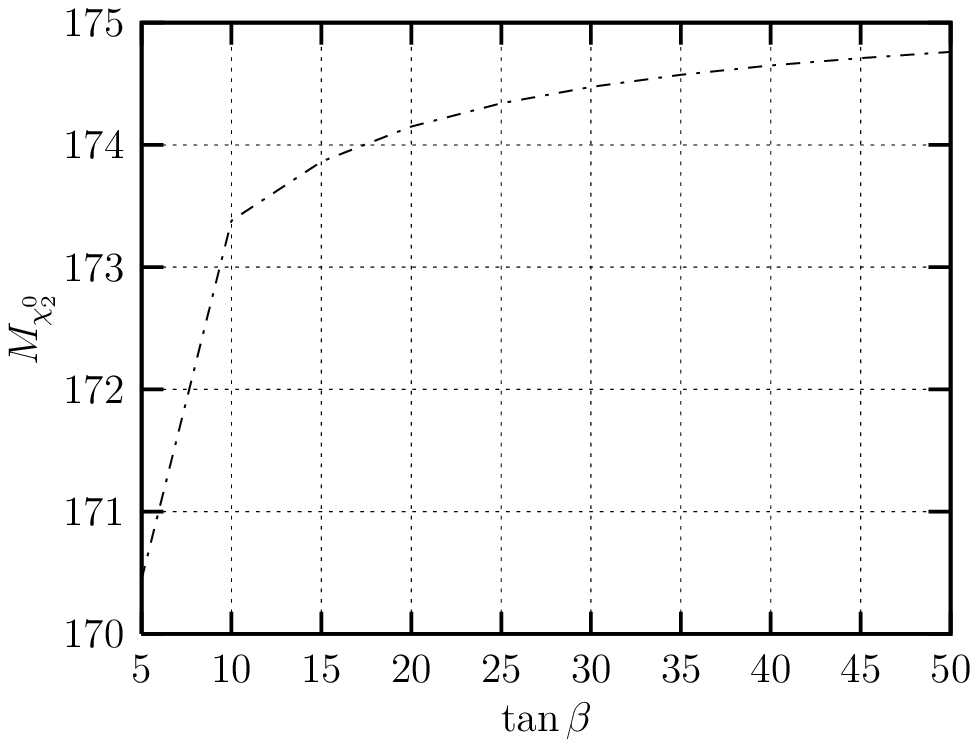,height=7.0in,width=5.0in }}
\vspace*{-2.9truein}
\fcaption{ The $\tan\beta$ dependence of 
$M_{{\chi}^{0}_{2}}$, when   
$M_{\chi^+_1}=105~\mbox{GeV}$~(left panel), and  
$M_{\chi^+_1}=160~\mbox{GeV}$~(right panel) $M_2>|\mu|$.} 
\label{fig4}
\end{figure}

In Figure 3 and in Figure 4, we plot the variation of the  second light 
neutralino mass $M_{{\chi}^{0}_{2}}$
with respect to $\tan\beta$,  
for $M_2 < |\mu|$  and  for $M_2 > |\mu|$ regimes, respectively.
In both Figures the left panels are for $M_{\chi^+_1}=105~\mbox{GeV}$, and  
the right panels are for $M_{\chi^+_1}=160~\mbox{GeV}$.

It can be seen from  Figure 3 and Figure 4 that  
similar to  the variation of $M_{{\chi}^{0}_{1}}$ (Figures 1 and 2),
$M_{{\chi}^{0}_{2}}$
increases as  $\tan\beta$ varies from 5 to 50, 
for both $M_2 < |\mu|$   and  $M_2 > |\mu|$ regimes.
The lower-upper bounds of $M_{{\chi}^{0}_{2}}$   
can be read as  90-104 $\mbox{GeV}$ when  $M_{\chi^+_1}=105~\mbox{GeV}$~(left panel of Fig. 3), 
and  142-160~$\mbox{GeV}$ when  $M_{\chi^+_1}=160~\mbox{GeV}$~(right panel of Fig. 3) for $M_2 < |\mu|$.
In passing to $M_2 > |\mu|$ regime, it is seen that the lower-upper bounds of $M_{{\chi}^{0}_{2}}$  
increase by an amount of $\sim \%10$ (Figure 4).
\begin{figure}[htb]
\vspace*{-2.5truein}
\hspace*{0.3truein}
\centerline{\psfig{file=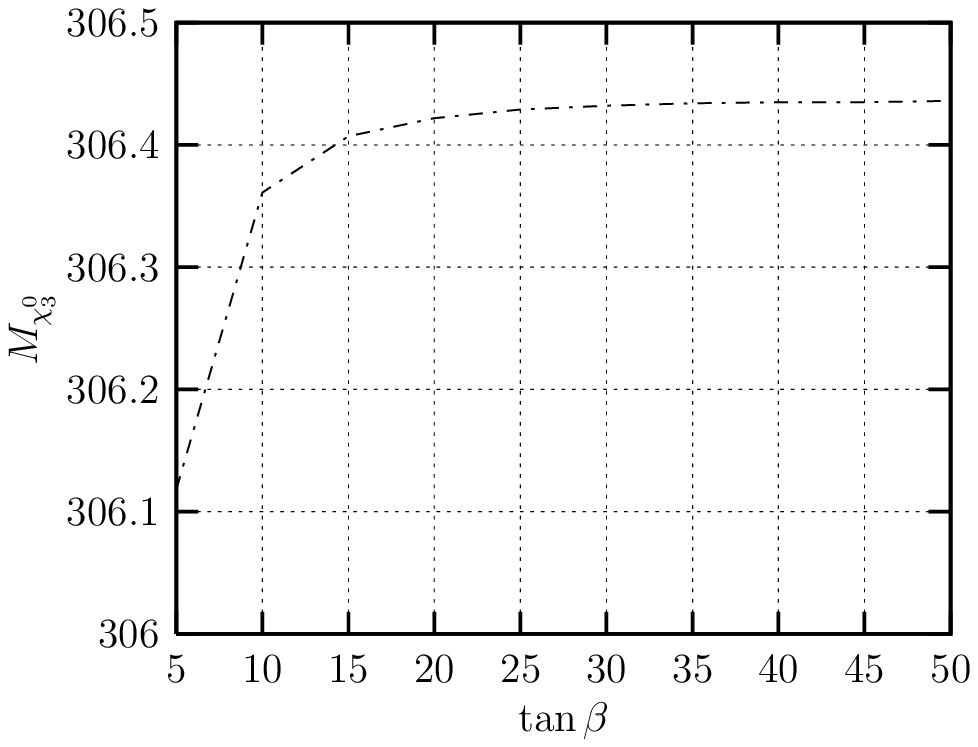,height=7.0in,width=5.0in }
\hspace*{-2.6truein}
\psfig{file=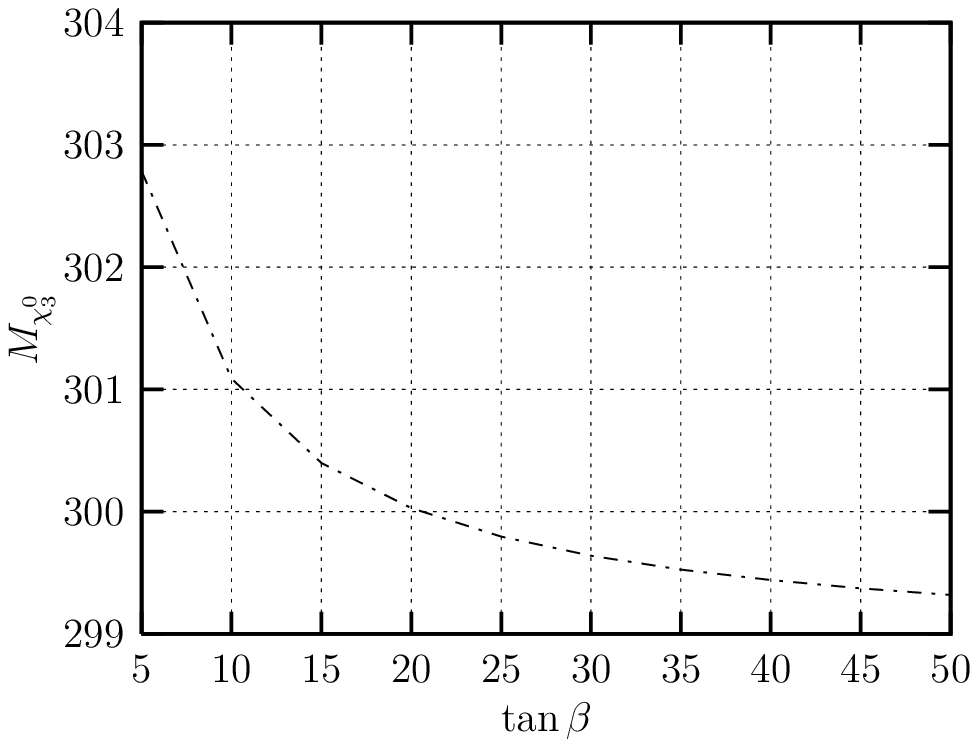,height=7.0in,width=5.0in }}
\vspace*{-2.9truein}
\fcaption{ The $\tan\beta$ dependence of 
$M_{{\chi}^{0}_{3}}$, when   
$M_{\chi^+_1}=105~\mbox{GeV}$~(left panel), and  
$M_{\chi^+_1}=160~\mbox{GeV}$~(right panel) for $M_2<|\mu|$.} 
\label{fig5}
\end{figure}
\begin{figure}[htb]
\vspace*{-2.5truein}
\hspace*{0.3truein}
\centerline{\psfig{file=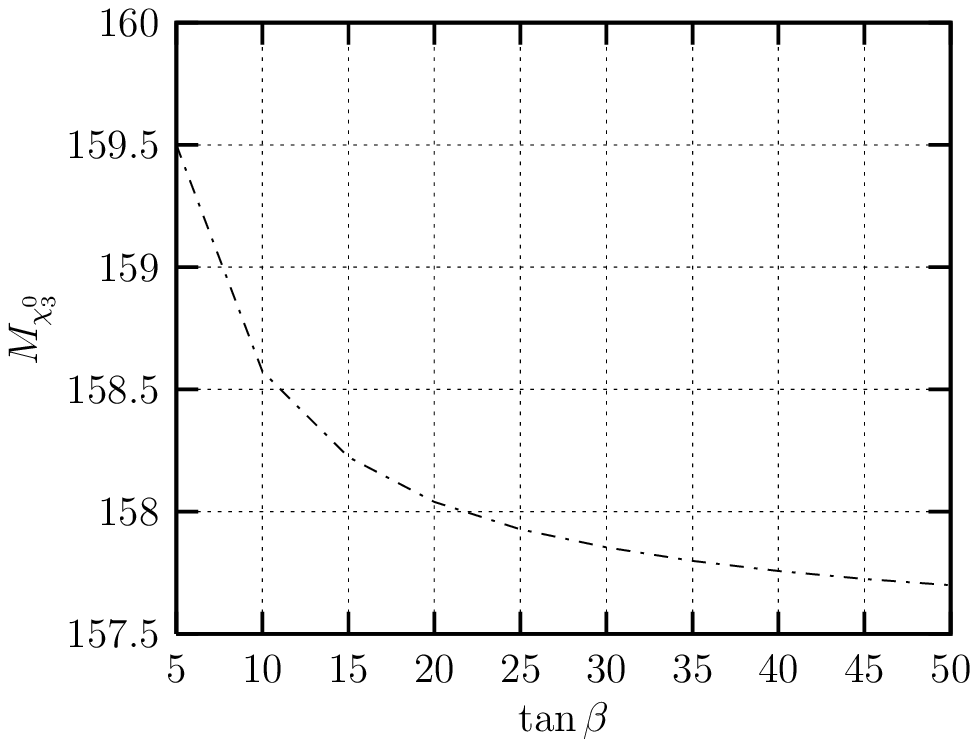,height=7.0in,width=5.0in }
\hspace*{-2.6truein}
\psfig{file=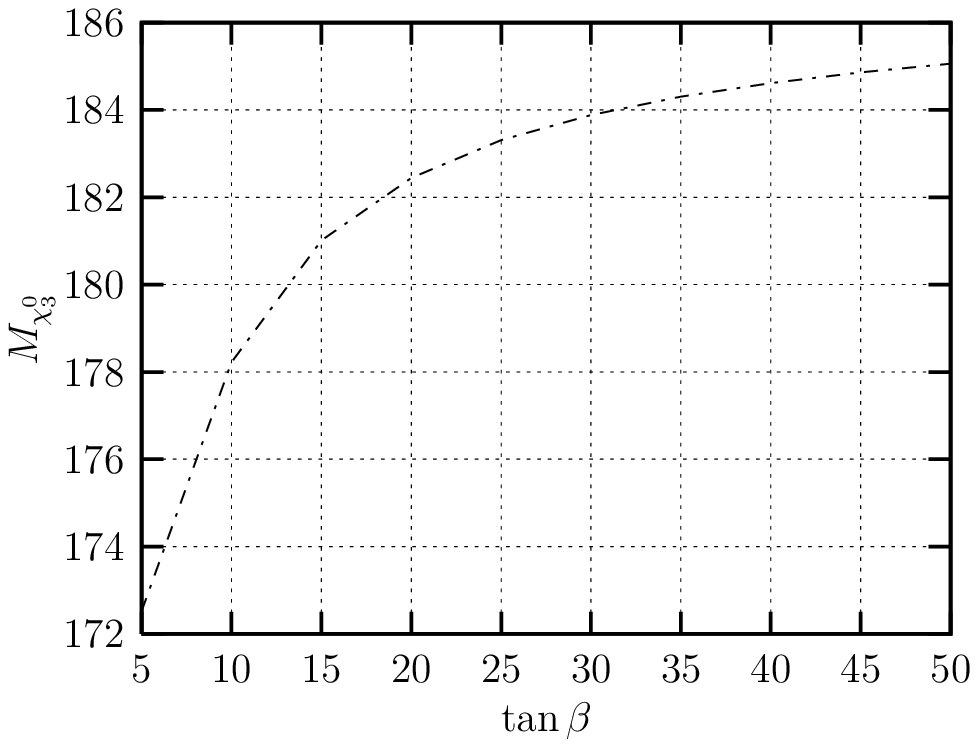,height=7.0in,width=5.0in }}
\vspace*{-2.9truein}
\fcaption{ The $\tan\beta$ dependence of 
$M_{{\chi}^{0}_{3}}$, when   
$M_{\chi^+_1}=105~\mbox{GeV}$~(left panel), and  
$M_{\chi^+_1}=160~\mbox{GeV}$~(right panel) for $M_2>|\mu|$.} 
\label{fig6}
\end{figure}

Up to now we have studied the $\tan\beta$ behaviour of the lighter neutralinos
($M_{{\chi}^{0}_{1}}$
and  $M_{{\chi}^{0}_{2}}$). The assigned values for the 
fundamental parameters in our numerical analysis indeed satisfy the assumption 
which went into the 
expressions (15)-(16). 

Next, we pass to the heavier neutralinos in  which case   
we plot the variation of the  next-to heaviest neutralino $M_{{\chi}^{0}_{3}}$
mass with respect to $\tan\beta$,  
for $M_2 < |\mu|$   and   for  $M_2 > |\mu|$ regimes in Figure 5 and in Figure 6, respectively.
In both of the Figures the left panels are for $M_{\chi^+_1}=105~\mbox{GeV}$, and  
the right panels are for $M_{\chi^+_1}=160~\mbox{GeV}$.

We observe from  Figure 5 that the 
behaviour of the ${\chi}^{0}_{3}$ mass with respect to   $\tan\beta$  
is the same with the lighter neutralinos
(${\chi}^{0}_{1}$ and ${\chi}^{0}_{2}$), for  the lighter  
chargino ($M_{\chi^+_1}=105~ \mbox{GeV}$),  however it 
reverses  for the heavier  chargino ($M_{\chi^+_1}=160~\mbox{GeV}$) for  the $M_2 < |\mu|$
regime. To understand this interesting behaviour it may be useful to look into the 
analytic expression (17), where  $M_{{\chi}^{0}_{3}}$ is related to  
the gaugino masses of $M_1$ and $M_2$ 
by the  combinations of 
$Z_1$ and $Z_2$. 
It can be observed that among the two contributions to the expression (17),
the first term of  (17) always dominates, as compared to the second term
for both    $M_{\chi^+_1}=105~\mbox{GeV}$ and  $M_{\chi^+_1}=160~\mbox{GeV}$,
cases.

One notes that this term increases with 
increasing $\tan\beta$ for 
the  lighter chargino mass ($M_{\chi^+_1}=105~\mbox{GeV}$), whereas
it decreases  
for the heavier chargino mass ($M_{\chi^+_1}=160~\mbox{GeV}$).
Since the contribution of this term is dominant,  the neutralino mass gets heavier with the 
increase in $\tan\beta$
for the lighter 
chargino mass (the left panel of Figure 5), and the behaviour is reversed for the heavier chargino mass
(the right panel of Figure 5).

As can be seen from (17), as the  lighter chargino mass  $M_{\chi^+_1}$  moves from a 
lower  value ($M_{\chi^+_1}=105~\mbox{GeV}$)
to a  higher one ($M_{\chi^+_1}=160~\mbox{GeV}$), then the three different
$\tan\beta$ contributions compete against each other, and their roles are 
changed at a certain critical value. 
This is the reason for the shift of the pattern of Figure 5 from one panel 
to the other. 
The critical value of the chargino mass is $M_{\chi^+_1}=130~\mbox{GeV}$
at which the $\tan\beta$-$M_{{\chi}^{0}_{3}}$ behaviour reverses. 

Similar observations can be made for the 
$M_2 > |\mu|$ regime,
by taking into account of the fact that 
the roles of $M_2$ and $|\mu|$ are interchanged for this case,
under the substitution: 
$M_2 \rightarrow |\mu|$, and $ \mu \rightarrow \mbox{sign} 
(\mu) \, M_2$~\cite{Choi3,Moultaka99}.

That this  behaviour is observed for  
the mass of ${\chi}^{0}_{3}$ particularly,
is  due to the fact that
its mass lies in the  transitional region from 
the lighter chargino masses to the  heavier.
\begin{figure}[htb]
\vspace*{-2.5truein}
\hspace*{0.3truein}
\centerline{\psfig{file=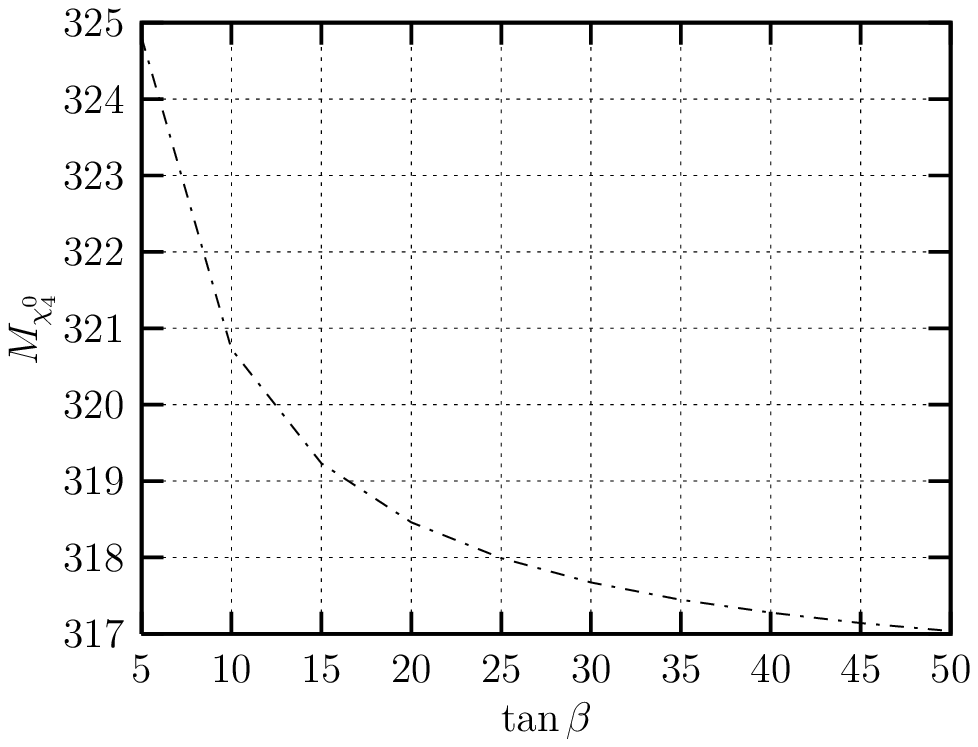,height=7.0in,width=5.0in }
\hspace*{-2.6truein}
\psfig{file=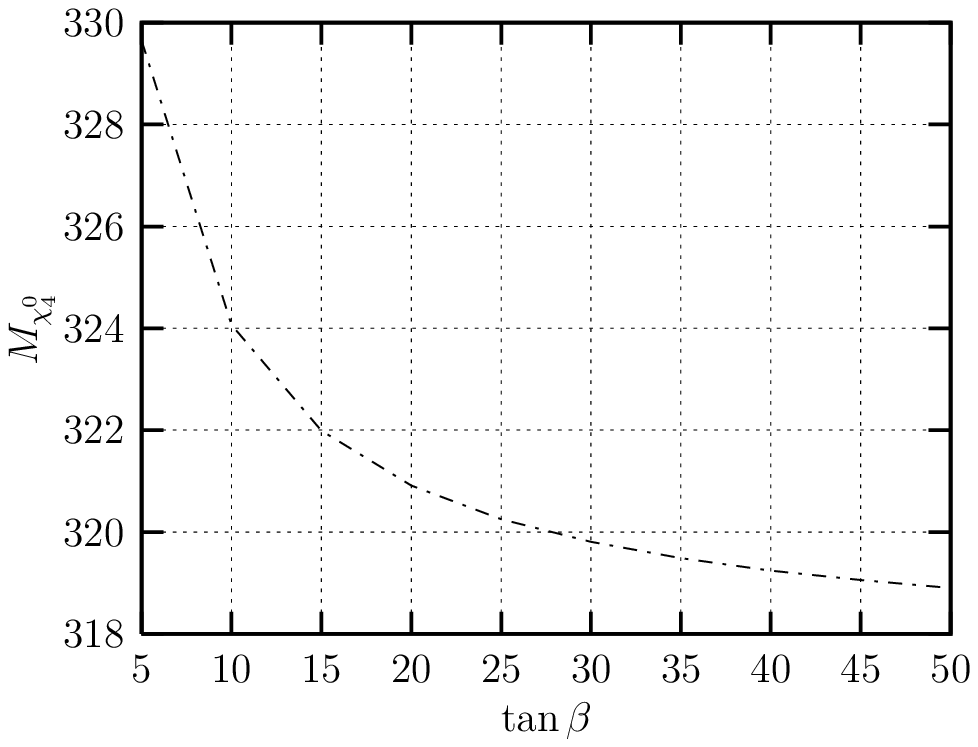,height=7.0in,width=5.0in }}
\vspace*{-2.9truein}
\fcaption{ The $\tan\beta$ dependence of 
$M_{{\chi}^{0}_{4}}$, when   
$M_{\chi^+_1}=105~\mbox{GeV}$~(left panel), and  
$M_{\chi^+_1}=160~\mbox{GeV}$~(right panel) for $M_2<|\mu|$.} 
\label{fig7}
\end{figure}
\begin{figure}[htb]
\vspace*{-2.5truein}
\hspace*{0.3truein}
\centerline{\psfig{file=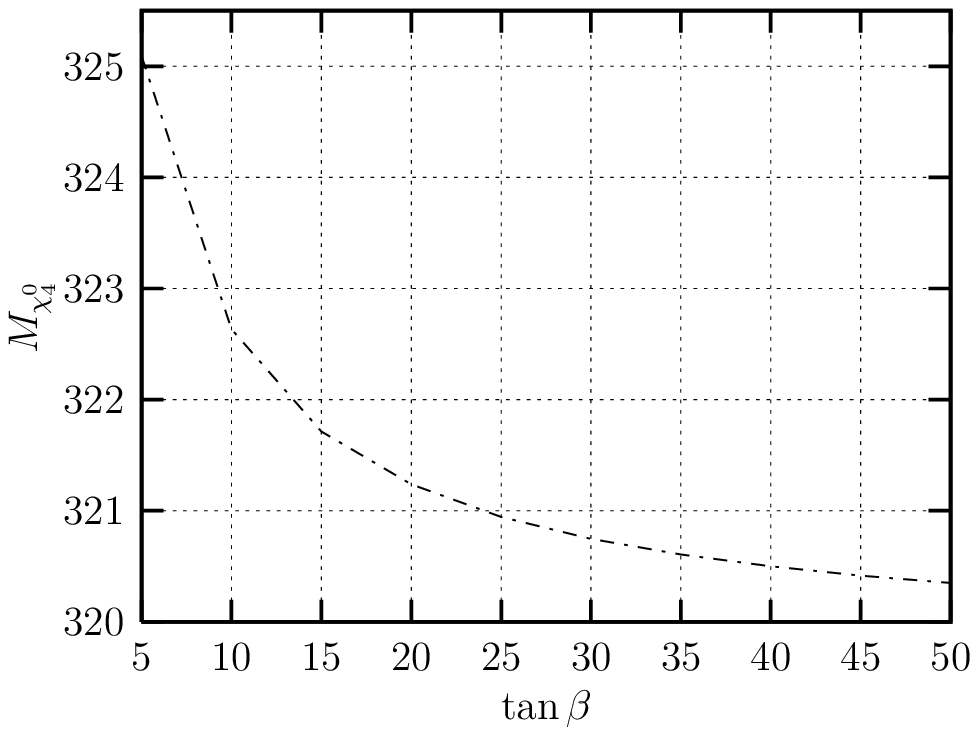,height=7.0in,width=5.0in }
\hspace*{-2.6truein}
\psfig{file=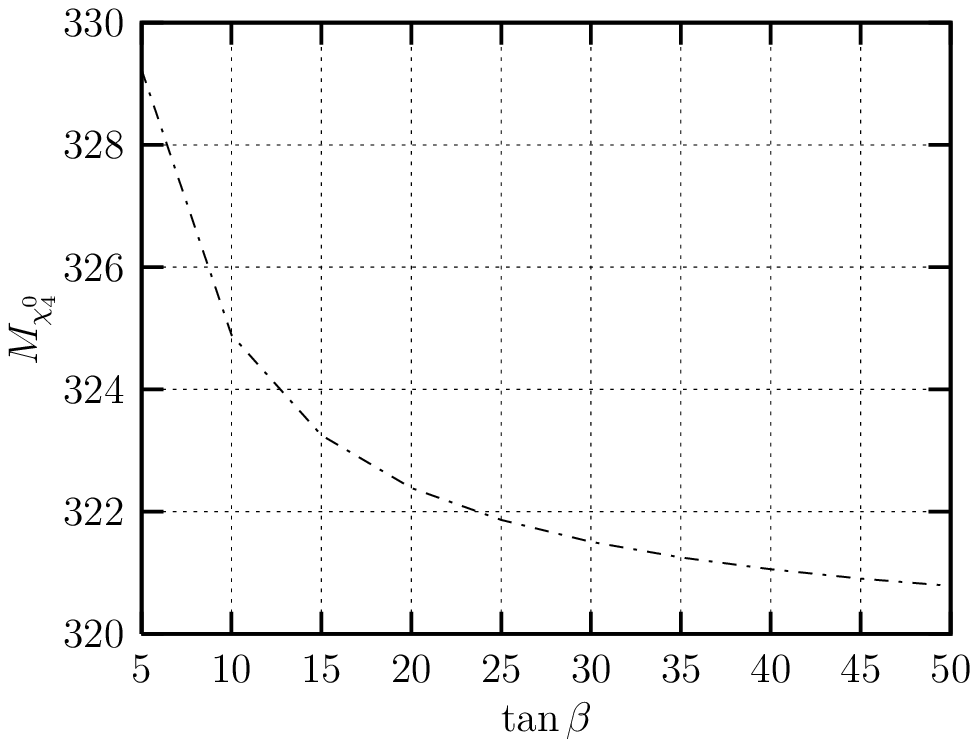,height=7.0in,width=5.0in }}
\vspace*{-2.9truein}
\fcaption{ The $\tan\beta$ dependence of 
$M_{{\chi}^{0}_{4}}$, when   
$M_{\chi^+_1}=105~\mbox{GeV}$~(left panel), and  
$M_{\chi^+_1}=160~\mbox{GeV}$~(right panel) for $M_2>|\mu|$.} 
\label{fig8}
\end{figure}

In Figure 7 and in Figure 8, we plot the variation of the mass of the heaviest 
neutralino $M_{{\chi}^{0}_{4}}$ with respect to $\tan\beta$,  
for $M_2 < |\mu|$ and for $M_2 > |\mu|$ regimes, respectively.
In both Figures the left panels are for $M_{\chi^+_1}=105~\mbox{GeV}$, and  
the right panels are for $M_{\chi^+_1}=160~\mbox{GeV}$.

We see from  Figure  7 and Figure 8 that   
$M_{{\chi}^{0}_{4}}$ decreases, as $\tan\beta$ increases.
Like $M_{{\chi}^{0}_{3}}$, it can be observed that among the two contributions to (18),
the first term of  (18) always dominates, as compared to the second term
for both    $M_{\chi^+_1}=105~\mbox{GeV}$ and  $M_{\chi^+_1}=160~\mbox{GeV}$,
cases. One notes that this term decreases 
with increasing $\tan\beta$
for both the  lighter  and the heavier chargino masses 
($M_{\chi^+_1}=105~\mbox{GeV}$ and $M_{\chi^+_1}=160~\mbox{GeV}$, respectively).
Since the contribution of this term is dominant,  $M_{{\chi}^{0}_{4}}$ gets lightened 
with the increase in $\tan\beta$.

A comparative analysis of  Figure 7 and Figure 8 suggest that  the mass of the heaviest neutralino    
remains around  $325~\mbox{GeV}$ for the lighter chargino  ($M_{\chi^+_1}=105~\mbox{GeV}$),
and does not exceed $330~\mbox{GeV}$,
for the heavier chargino  ($M_{\chi^+_1}=160~\mbox{GeV}$).

\subsection{CP violating case}

In the second part of our analysis, we carry out the analysis when there is CP 
violation. In the following, we analyze the $\varphi_{\mu}$ dependence of 
$M_{{\chi}^{0}_{i}}$, as   $\varphi_{\mu}$  ranges from 0 to $ 2 \pi$.
In our analysis, we fix $M_{\chi^+_2}=320~\mbox{GeV}$ and 
we choose two values of the lightest chargino mass, like the CP conserving case.
Here, we consider two specific values of
$\tan\beta $; Namely,  $\tan\beta=5$, and  $\tan\beta=50$, representing the low and
high $\tan\beta$ regimes. 
\begin{figure}[htb]
\vspace*{-2.5truein}
\hspace*{0.3truein}
\centerline{\psfig{file=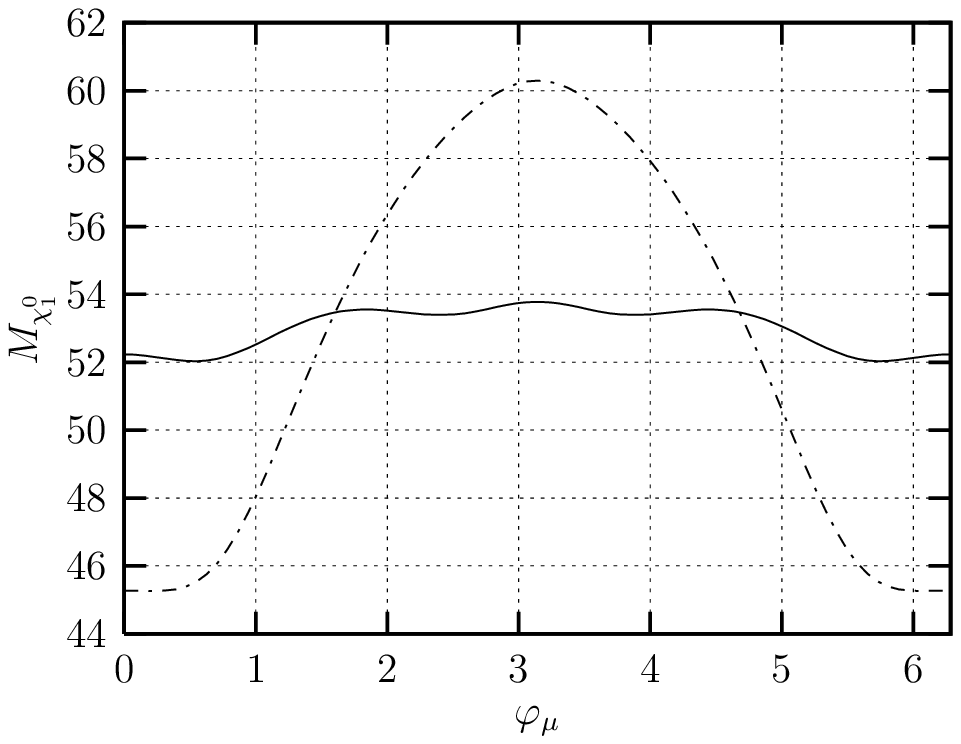,height=7.0in,width=5.0in }
\hspace*{-2.6truein}
\psfig{file=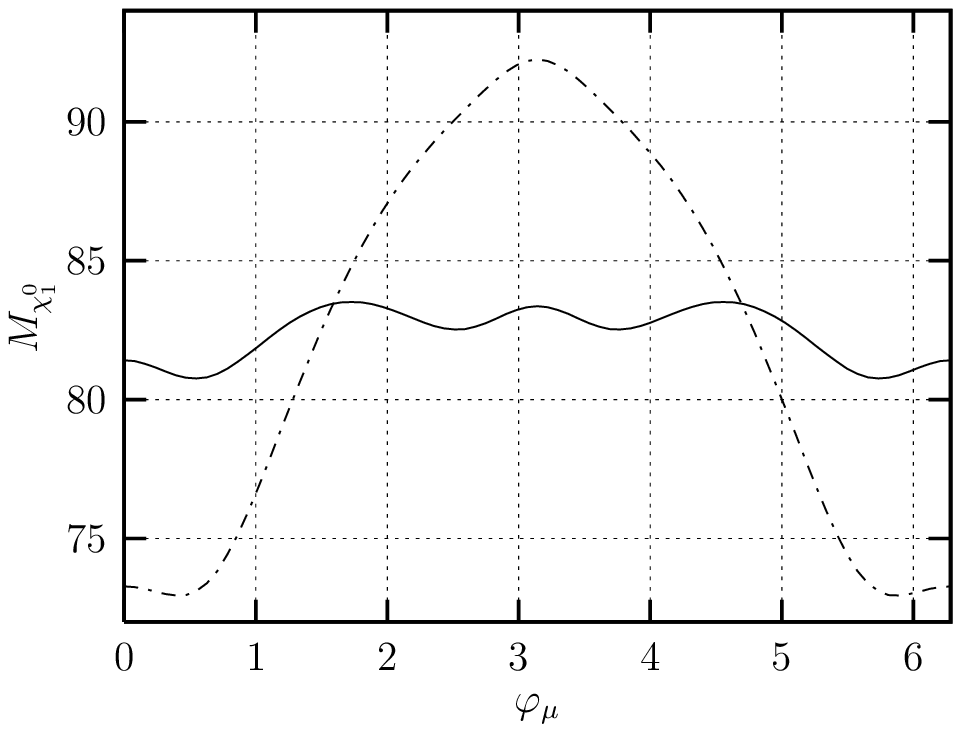,height=7.0in,width=5.0in }}
\vspace*{-2.9truein}
\fcaption{The $\varphi_{\mu}$ dependence of  
$M_{{\chi}^{0}_{1}}$, when   $M_{\chi^+_1}=105~\mbox{GeV}$~(left panel),
and $M_{\chi^+_1}=160~\mbox{GeV}$~(right panel), for $M_2<|\mu|$. } 
\label{fig9}
\end{figure}
\begin{figure}[htb]
\vspace*{-2.5truein}
\hspace*{0.3truein}
\centerline{\psfig{file=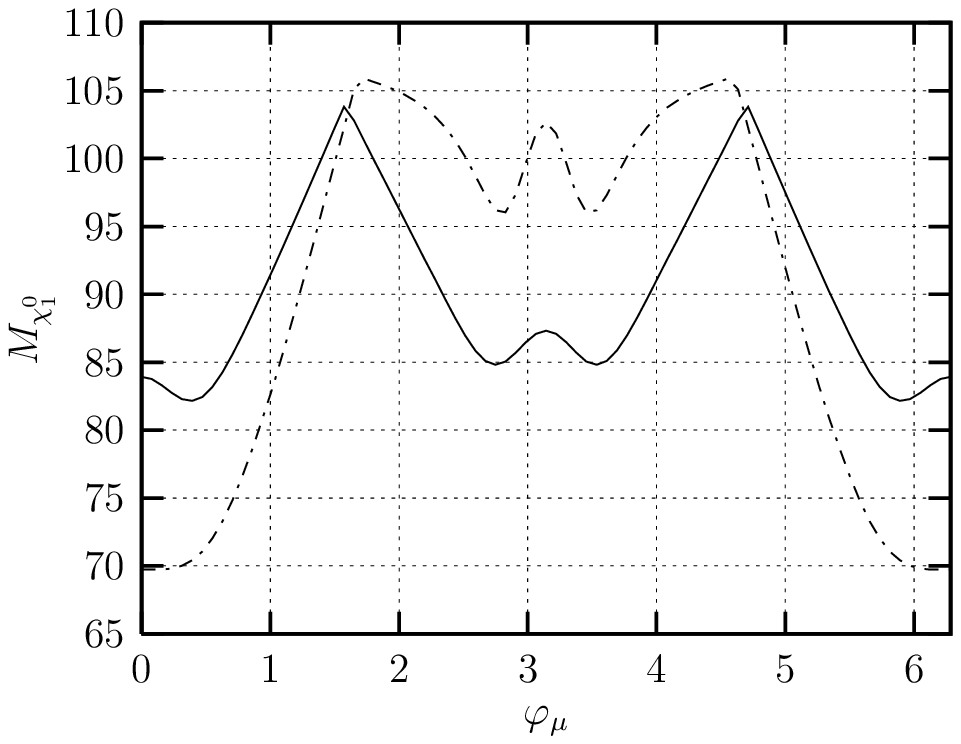,height=7.0in,width=5.0in }
\hspace*{-2.6truein}
\psfig{file=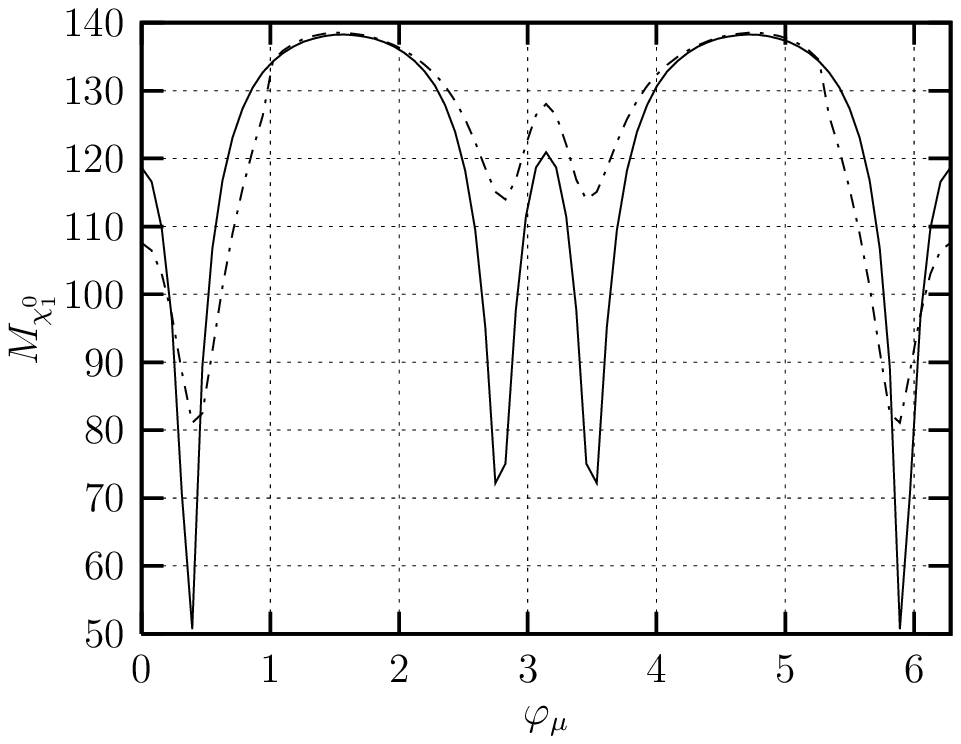,height=7.0in,width=5.0in }}
\vspace*{-2.9truein}
\fcaption{The $\varphi_{\mu}$ dependence of  
$M_{{\chi}^{0}_{1}}$, when   $M_{\chi^+_1}=105~\mbox{GeV}$~(left panel),
and $M_{\chi^+_1}=160~\mbox{GeV}$~(right panel), for $M_2>|\mu|$. } 
\label{fig10}
\end{figure}

In Figures 9, and 10, we show  the $\varphi_{\mu}$ dependence of  
the mass of the lightest neutralino 
$M_{{\chi}^{0}_{1}}$, at  $M_{\chi^+_1}=105~\mbox{GeV}$~(left panels), and  
$M_{\chi^+_1}=160~\mbox{GeV}$~(right panels), for  $M_2 < |\mu|$,
and  $M_2 > |\mu|$ regimes, respectively. In 
each panel the dotted  curves are for  $\tan\beta=5$, whereas
the solid  ones are for   $\tan\beta=50$. 

When $M_{\chi^+_1}=105~\mbox{GeV}$, 
$M_{{\chi}^{0}_{1}}$ changes from  45 to 60~$\mbox{GeV}$
and  from 70 to 105~$\mbox{GeV}$, for the  $M_2 < |\mu|$ and  $M_2 >|\mu|$ regimes, respectively, at $\tan\beta=5$,
as  $\varphi_{\mu}$ varies in the  [0 , $ \pi$] interval.
For $M_{\chi^+_1}=160~\mbox{GeV}$,  it is 
seen that the lower 
and upper bounds of $M_{{\chi}^{0}_{1}}$
increases for both  $M_2 <|\mu|$ and $M_2> |\mu|$ regimes, as expected.

A comparative analysis of Figures 9 and Figure 10  
suggest that  $M_{{\chi}^{0}_{1}}$  is quite
sensitive to the variations of $\varphi_{\mu}$.
It is interesting to note  that  
when  $M_{\chi^+_1}=105~\mbox{GeV}$,
$\varphi_{\mu}$  dependence of  
$M_{{\chi}^{0}_{1}}$ is  sharper at the CP violating points ($\varphi_{\mu}$=$\pi/2$  and  $3\pi/2$), 
for  the   $M_2 >|\mu|$  regime, at both  $\tan\beta=5$ and  $\tan\beta=50$.
When $M_{\chi^+_1}=160~\mbox{GeV}$~(right panel of Figure 10), 
one observes a slower variation at the CP violating points.
On the other hand, for the   $M_2< |\mu|$ regime, it can be seen that
the variation of  $\varphi_{\mu}$ 
(at both $M_{\chi^+_1}=105~\mbox{GeV}$ and 
$M_{\chi^+_1}=160~\mbox{GeV}$),
is slower  as compared to  $M_2 >|\mu|$  regime (see, for instance the analytical
expression of  $M_{{\chi}^{0}_{1}}$ given by (15)).

One notes that, when   $M_{\chi^+_2}=320~\mbox{GeV}$, 
$(i)$ $|\mu|$ changes from $299~\mbox{GeV}$, to $292~\mbox{GeV}$,
whereas $M_2$  from $104~\mbox{GeV}$ to  $124~\mbox{GeV}$, for
$M_{\chi^+_1}=105~\mbox{GeV}$, 
$(ii)$ $|\mu|$ changes from $297~\mbox{GeV}$, to $280~\mbox{GeV}$,
whereas $M_2$  from $164~\mbox{GeV}$ to  $192~\mbox{GeV}$, for
$M_{\chi^+_1}=160~\mbox{GeV}$ at $\tan\beta=5$.  
On the other hand, 
$(i)$ $|\mu|$ changes from $296~\mbox{GeV}$, to $295~\mbox{GeV}$,
whereas $M_2$  from $113~\mbox{GeV}$ to  $115~\mbox{GeV}$, for
$M_{\chi^+_1}=105~\mbox{GeV}$, 
$(ii)$ $|\mu|$ changes from $290~\mbox{GeV}$, to $288~\mbox{GeV}$,
whereas $M_2$  from $175~\mbox{GeV}$ to  $178~\mbox{GeV}$, for
$M_{\chi^+_1}=160~\mbox{GeV}$ at $\tan\beta=50$.  
\begin{figure}[htb]
\vspace*{-2.5truein}
\hspace*{0.3truein}
\centerline{\psfig{file=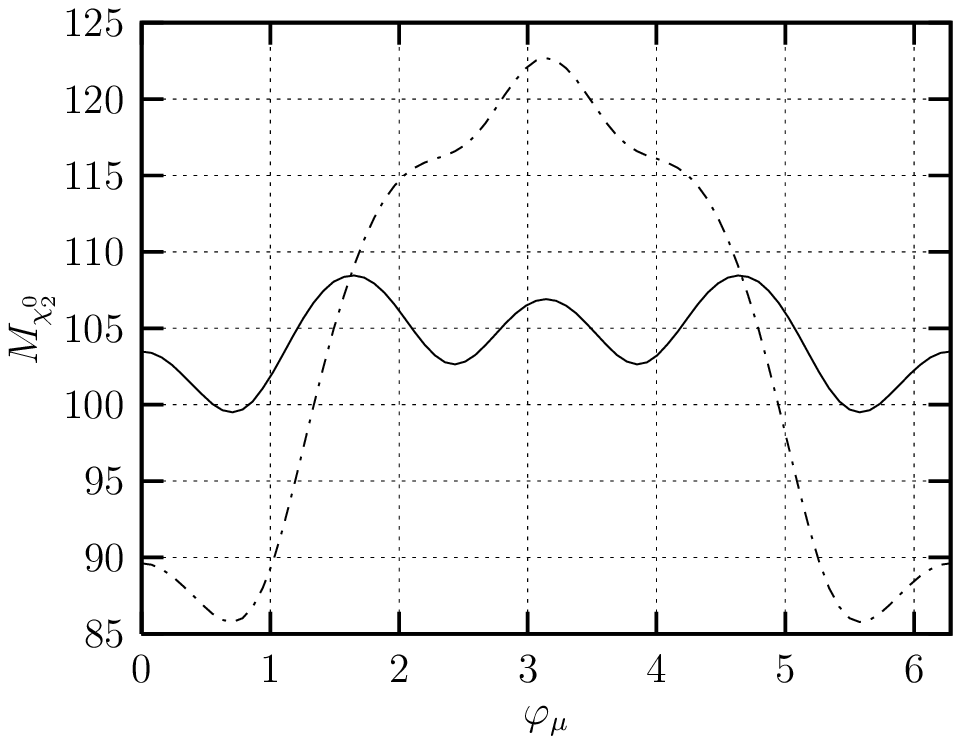,height=7.0in,width=5.0in }
\hspace*{-2.6truein}
\psfig{file=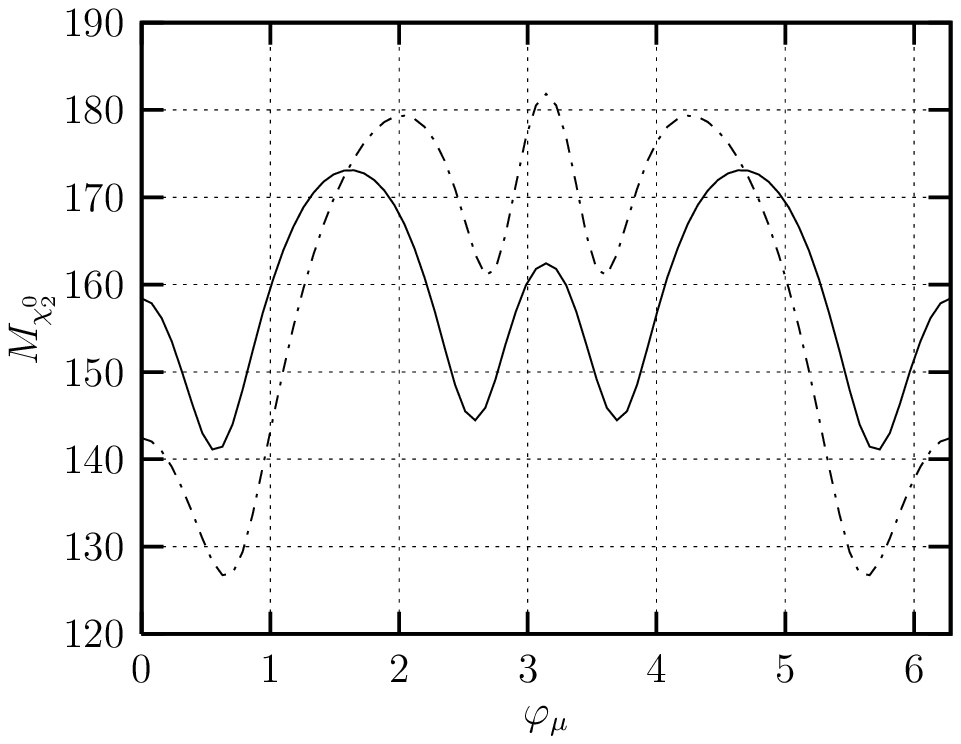,height=7.0in,width=5.0in }}
\vspace*{-2.9truein}
\fcaption{The $\varphi_{\mu}$ dependence of  
$M_{{\chi}^{0}_{2}}$, when   $M_{\chi^+_1}=105~\mbox{GeV}$~(left panel),
and $M_{\chi^+_1}=160~\mbox{GeV}$~(right panel), for $M_2<|\mu|$. } 
\label{fig11}
\end{figure}
\begin{figure}[htb]
\vspace*{-2.5truein}
\hspace*{0.3truein}
\centerline{\psfig{file=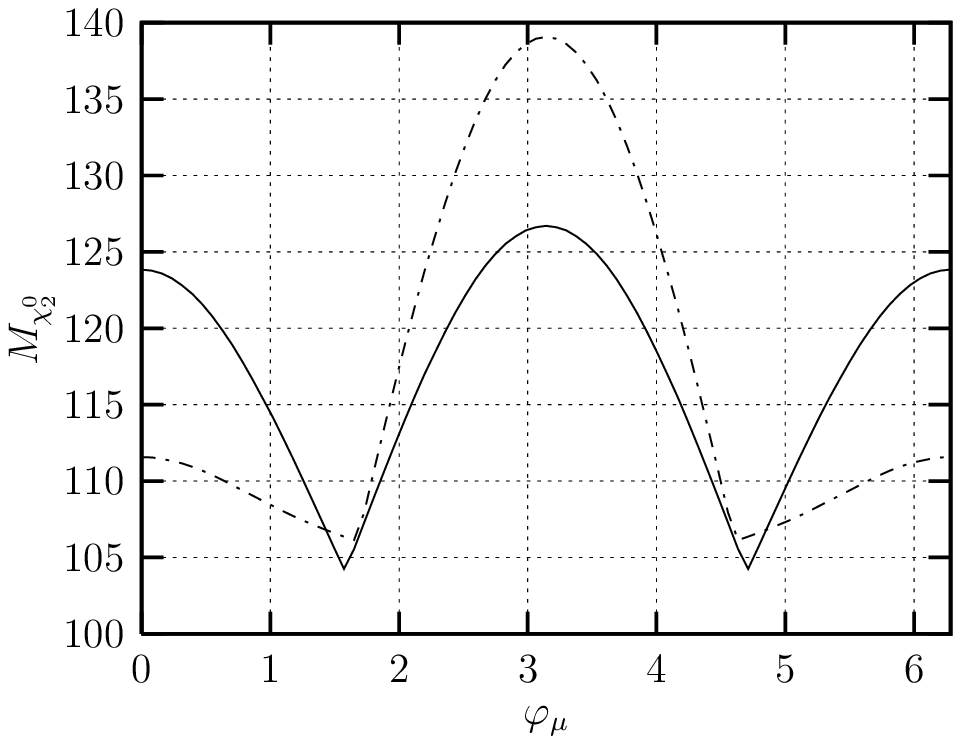,height=7.0in,width=5.0in }
\hspace*{-2.6truein}
\psfig{file=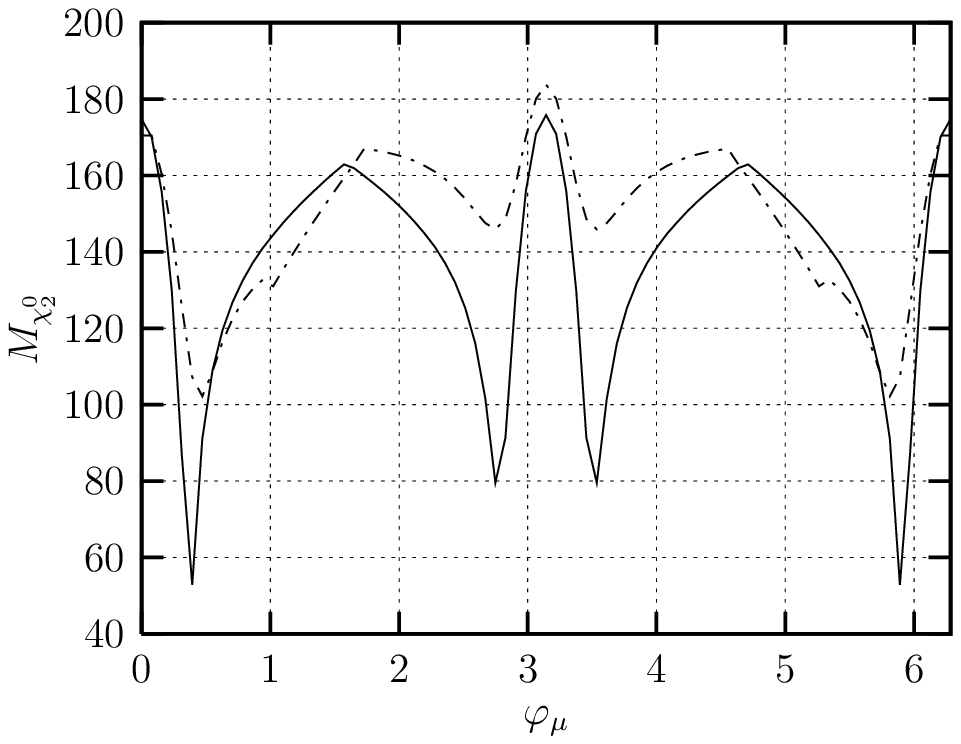,height=7.0in,width=5.0in }}
\vspace*{-2.9truein}
\fcaption{The $\varphi_{\mu}$ dependence of  
$M_{{\chi}^{0}_{2}}$, when   $M_{\chi^+_1}=105~\mbox{GeV}$~(left panel),
and $M_{\chi^+_1}=160~\mbox{GeV}$~(right panel), for $M_2>|\mu|$. } 
\label{fig12}
\end{figure}

In Figures 11, and 12, we show  the $\varphi_{\mu}$ dependence of  
$M_{{\chi}^{0}_{2}}$,  for  $M_2 < |\mu|$,
and  $M_2 > |\mu|$ regimes, respectively,
when  $M_{\chi^+_1}=105~\mbox{GeV}$~(left panels), and  
$M_{\chi^+_1}=160~\mbox{GeV}$~(right panels).
In each Figure, the dotted curves correspond to $\tan\beta=5$, whereas 
the solid curves to   $\tan\beta=50$. 

One notes from the left panel of Figure 11 that 
the variation is $M_{{\chi}^{0}_{2}}$,
is more faster as compared to $M_{{\chi}^{0}_{1}}$ (see Figure 9).
Such  behaviour can be explained by referring into the analytic expression 
of $M_{{\chi}^{0}_{2}}$ (16).

On the other hand,
as can be seen from the left panel of Figure 12,  starting from  $\varphi_{\mu}=0$
at  $112~\mbox{GeV}$, $M_{{\chi}^{0}_{2}}$
decreases to $105~\mbox{GeV}$  at $\varphi_{\mu}=\pi/2$,
then it increases to   $140~\mbox{GeV}$ at  $\varphi_{\mu}=\pi$,
at $\tan\beta=5$. For the heavier chargino~(right panel)
one observes a faster and sharper variation of  $\varphi_{\mu}$,
as compared to the lighter chargino~(left panel).
However, it is seen that $M_{{\chi}^{0}_{2}}$
gets heavier together with the heavier chargino mass ($M_{\chi^+_1}=160~ \mbox{GeV}$),
without causing too big splitting among the low and high $\tan\beta$
regimes. For instance, the maximal values at   $\varphi_{\mu}=\pi$, for  $\tan\beta=5$  and 
$\tan\beta=50$ cases,  are  $178~\mbox{GeV}$ and $180~\mbox{GeV}$, respectively.
\begin{figure}[htb]
\vspace*{-2.5truein}
\hspace*{0.3truein}
\centerline{\psfig{file=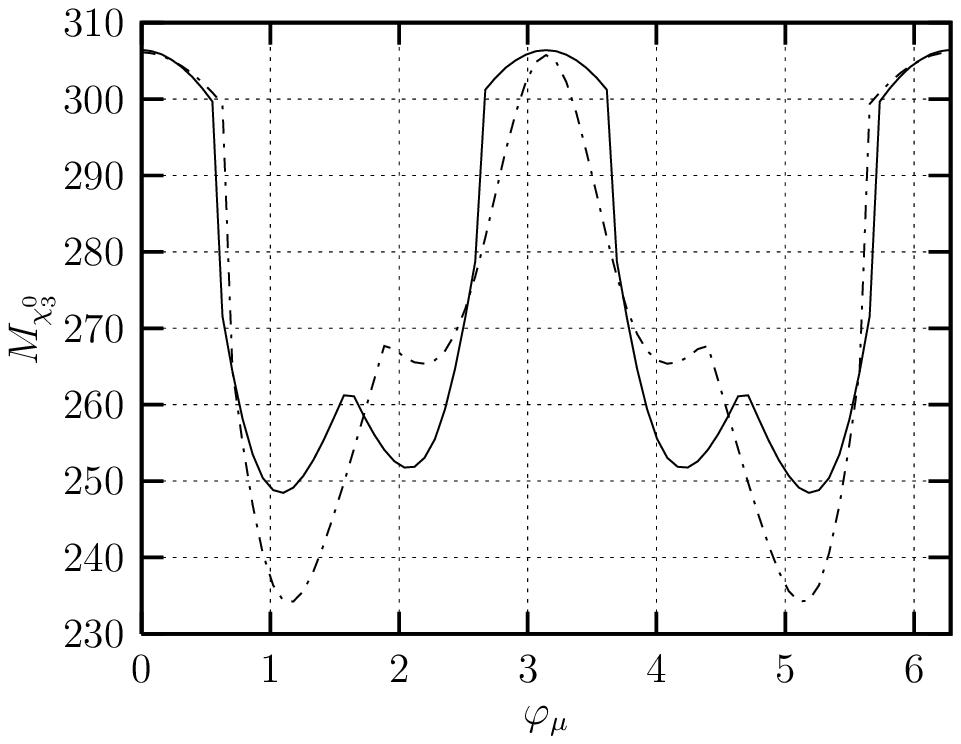,height=7.0in,width=5.0in }
\hspace*{-2.6truein}
\psfig{file=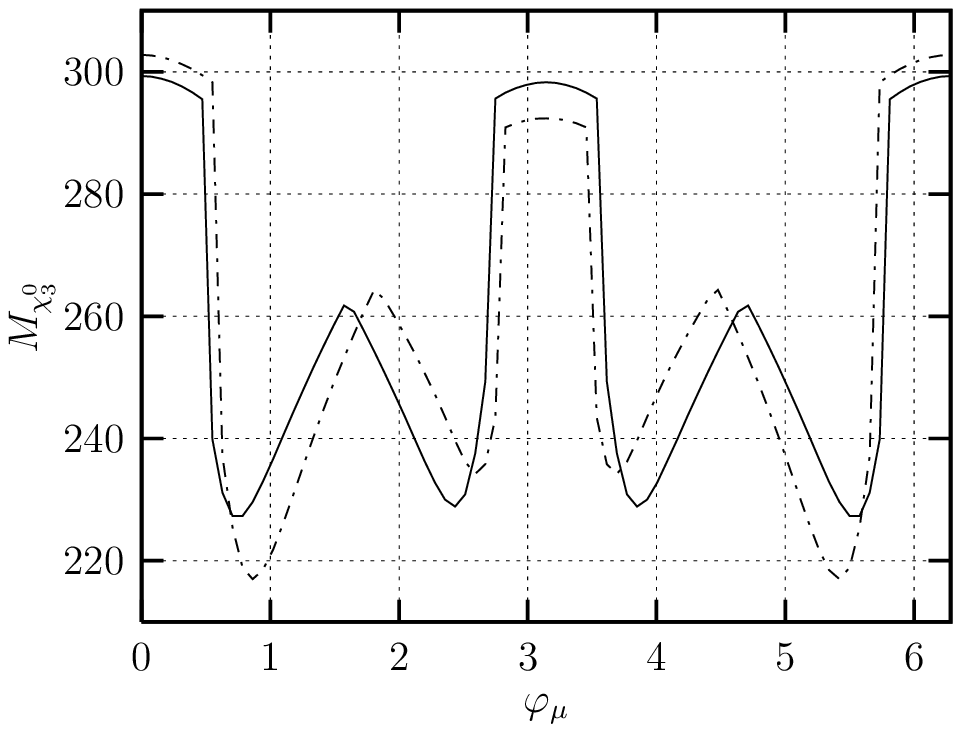,height=7.0in,width=5.0in }}
\vspace*{-2.9truein}
\fcaption{The $\varphi_{\mu}$ dependence of  
$M_{{\chi}^{0}_{3}}$, when   $M_{\chi^+_1}=105~\mbox{GeV}$~(left panel),
and $M_{\chi^+_1}=160~\mbox{GeV}$~(right panel), for $M_2<|\mu|$. } 
\label{fig13}
\end{figure}
\begin{figure}[htb]
\vspace*{-2.5truein}
\hspace*{0.3truein}
\centerline{\psfig{file=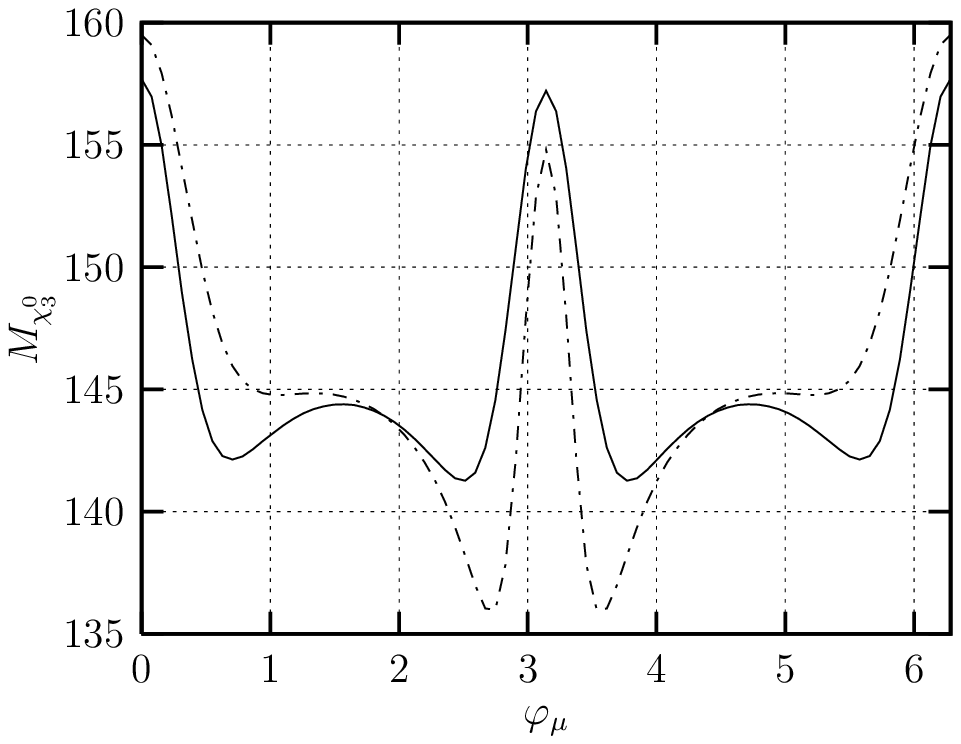,height=7.0in,width=5.0in }
\hspace*{-2.6truein}
\psfig{file=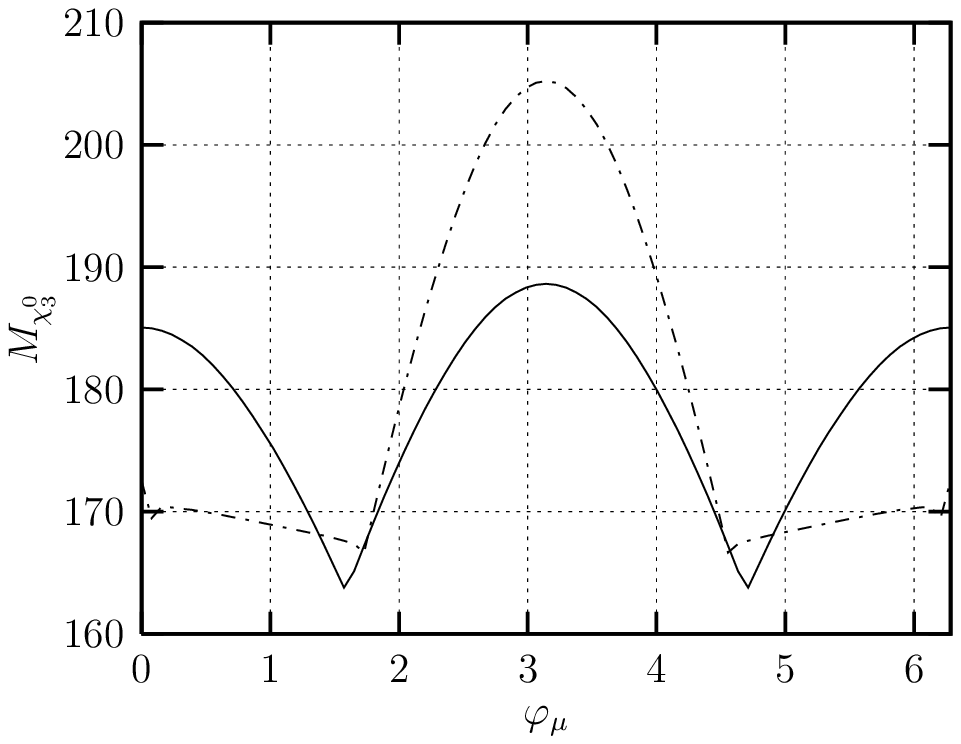,height=7.0in,width=5.0in }}
\vspace*{-2.9truein}
\fcaption{The $\varphi_{\mu}$ dependence of  
$M_{{\chi}^{0}_{3}}$, when   $M_{\chi^+_1}=105~\mbox{GeV}$~(left panel),
and $M_{\chi^+_1}=160~\mbox{GeV}$~(right panel), for $M_2>|\mu|$. } 
\label{fig14}
\end{figure}

In Figures 13, and 14, we show  the $\varphi_{\mu}$ dependence of  
$M_{{\chi}^{0}_{3}}$, at  $M_{\chi^+_1}=105~\mbox{GeV}$~(left panels), and  
$M_{\chi^+_1}=160~\mbox{GeV}$~(right panels), for  $M_2 < |\mu|$,
and  $M_2 > |\mu|$ regimes, respectively.
In the each panel,  the dotted curves correspond to $\tan\beta=5$, whereas 
the solid curves to   $\tan\beta=50$. 

It is seen from the left panel of Figure 13 that the variation of $M_{{\chi}^{0}_{3}}$
with respect to  $\varphi_{\mu}$ is much  faster as compared to the lighter 
neutralinos $M_{{\chi}^{0}_{1}}$, and $M_{{\chi}^{0}_{2}}$, as expected (see for 
instance the analytical expression of $M_{{\chi}^{0}_{3}}$ given by the expression 
(17)).  

It is interesting to note that 
$M_{{\chi}^{0}_{3}}$ at lighter  
$M_{\chi^+_1}$~($M_{\chi^+_1}=105~\mbox{GeV}$, the left panel
of Figure 12) has similar  $\varphi_{\mu}$ dependence
with   that of  $M_{{\chi}^{0}_{3}}$
at  heavier   $M_{\chi^+_1}$~($M_{\chi^+_1}=160~\mbox{GeV}$, the right 
panel of Figure 14).

Finally, in Figures 15, and 16, we show  the $\varphi_{\mu}$ dependence of  
$M_{{\chi}^{0}_{4}}$, at  $M_{\chi^+_1}=105~\mbox{GeV}$~(left panels), and  
$M_{\chi^+_1}=160~\mbox{GeV}$~(right panels), for  $M_2 < |\mu|$,
and  $M_2 > |\mu|$ regimes, respectively.
In the each panel,  the dotted curves correspond to $\tan\beta=5$, whereas 
the solid curves to   $\tan\beta=50$.
\begin{figure}[htb]
\vspace*{-2.5truein}
\hspace*{0.3truein}
\centerline{\psfig{file=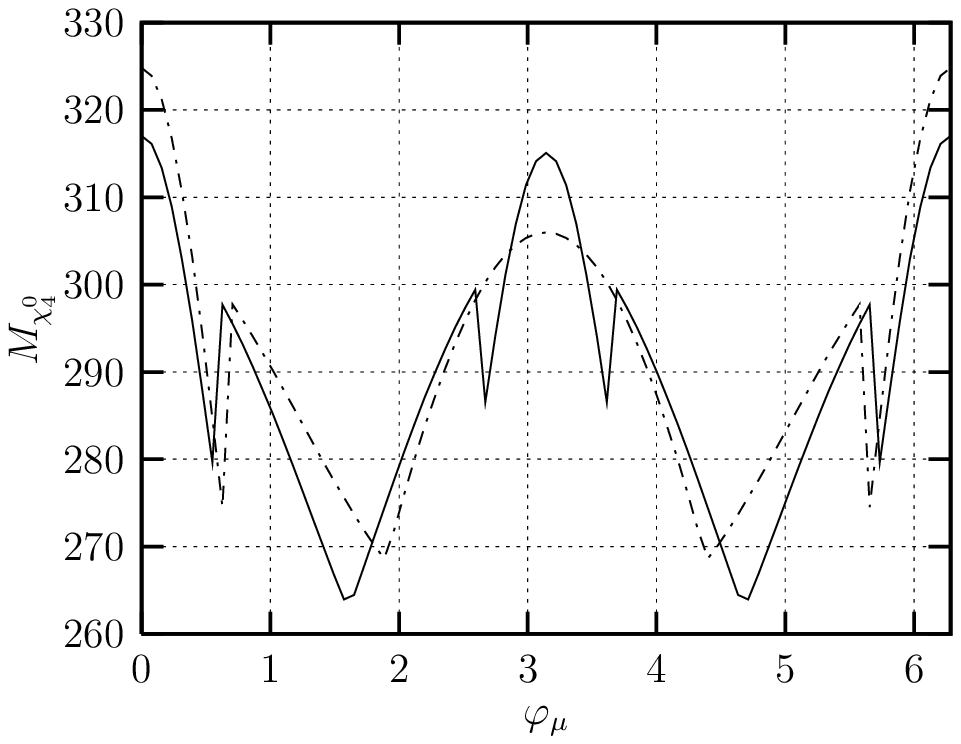,height=7.0in,width=5.0in }
\hspace*{-2.6truein}
\psfig{file=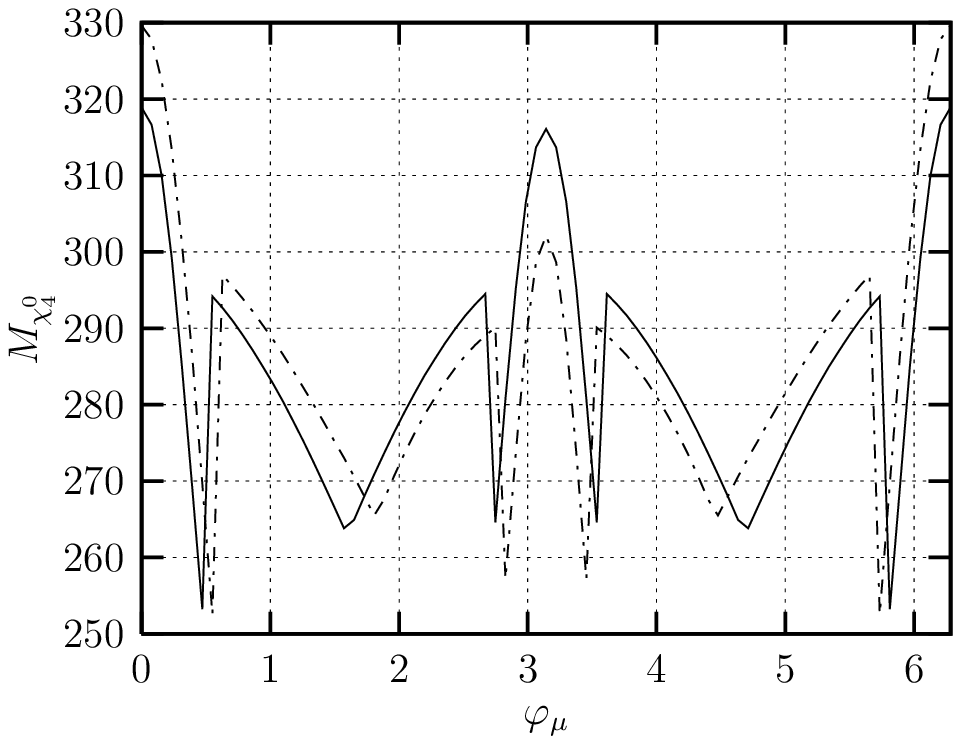,height=7.0in,width=5.0in }}
\vspace*{-2.9truein}
\fcaption{The $\varphi_{\mu}$ dependence of  
$M_{{\chi}^{0}_{4}}$, when   $M_{\chi^+_1}=105~\mbox{GeV}$~(left panel),
and $M_{\chi^+_1}=160~\mbox{GeV}$~(right panel), for $M_2<|\mu|$. } 
\label{fig15}
\end{figure}
\begin{figure}[htb]
\vspace*{-2.5truein}
\hspace*{0.3truein}
\centerline{\psfig{file=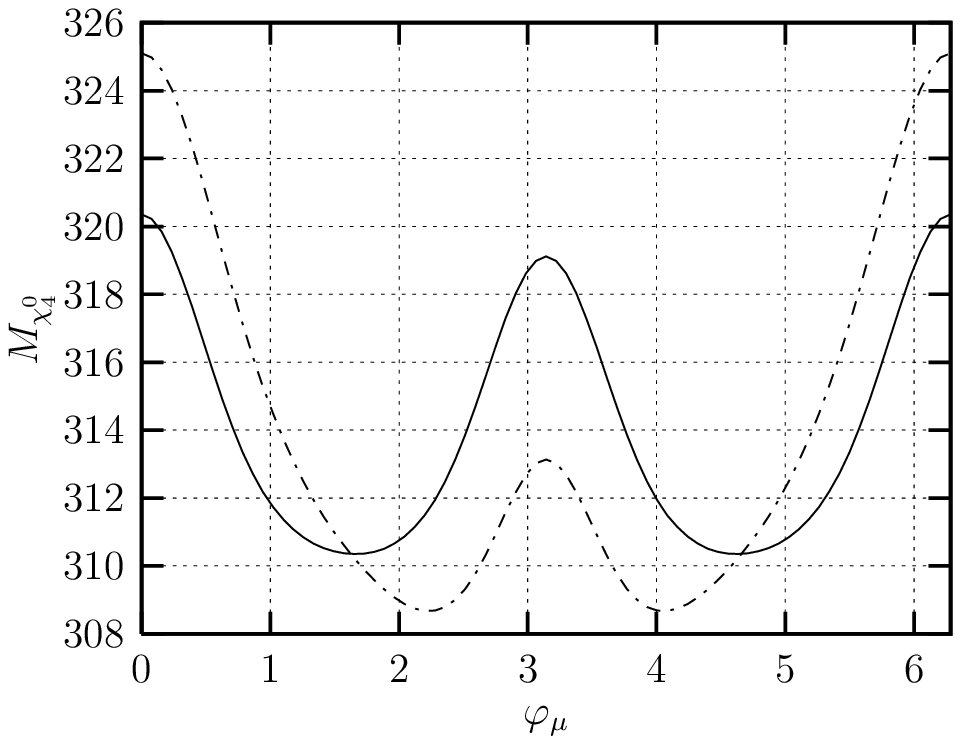,height=7.0in,width=5.0in }
\hspace*{-2.6truein}
\psfig{file=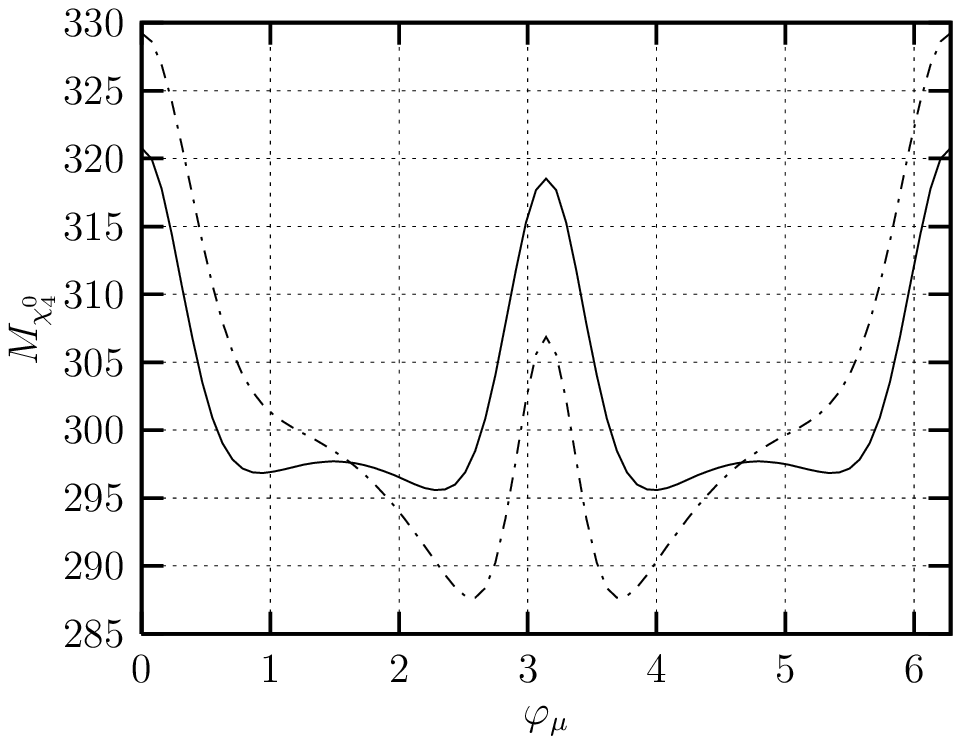,height=7.0in,width=5.0in }}
\vspace*{-2.9truein}
\fcaption{The $\varphi_{\mu}$ dependence of  
$M_{{\chi}^{0}_{4}}$, when   $M_{\chi^+_1}=105~\mbox{GeV}$~(left panel),
and $M_{\chi^+_1}=160~\mbox{GeV}$~(right panel), for $M_2>|\mu|$. } 
\label{fig16}
\end{figure}

A comparative analysis of Figures 13-16 suggest that  
for the heavier neutralinos (i=3,4), the dependence of the masses to $\varphi_{\mu}$ 
show complementary behaviour. 
For instance the CP violating points 
($\varphi_{\mu}=\pi/2, 3 \pi/2$) 
appear as local maxima 
for the ${\chi}^{0}_{3}$ mass,
as opposed to 
${\chi}^{0}_{4}$ case where
those values are local minima, for both values of  
$M_{\chi^+_1}=105~\mbox{GeV}$ and  
$M_{\chi^+_1}=160~\mbox{GeV}$,
in the  $M_2 < |\mu|$ regime. This 
complementarity at the CP violating points for the heavier sector can easily
be seen from the expressions (17)-(18).
A similar complementarity holds for  
the lighter neutralinos (i=1,2),
at the lighter chargino mass in the  $M_2 > |\mu|$ regime
(the left panels of Figures 10 and 12).
On the other hand, the CP violating points appear as local maxima
for  both values of the
lightest chargino masses in the $M_2 < |\mu|$ regime (Figures 9 and 11).

It can also be seen from Figures 9-16, that the variations of the lighter 
neutralino masses (i=1,2) with $\tan\beta$, is  about $\%15$ 
in the range from 5 to 50 at the CP violating points as in the CP conserving case.
However, for the heavier neutralino case, the difference of the masses  
between high and low $\tan\beta$ regimes becomes very small.

\section{Conclusions}

We have analyzed the neutralino system, whose parameters are
extracted from the chargino system, for both CP conserving and CP violating
cases. Here is a brief summary of our main results:

When $\varphi_{\mu}=0$,
given $M_{\chi^+_1}$,  $M_{\chi^+_2}$ and $\tan\beta$, the masses of all the 
neutralinos can be determined.
The variation of $\tan\beta$ from 5 to 50 leads to  at most 
$\%15$ change in the neutralino masses.

The  $\tan\beta$ behaviour of the lighter neutralinos  $M_{{\chi}^{0}_{i}}$
(i=1,2) are opposite to the that of the  heaviest
neutralino (i=4) for the lower and the higher values of the 
lighter chargino mass  ($M_{\chi^+_1}=105~\mbox{GeV}$, and $M_{\chi^+_1}=160~\mbox{GeV}$),
for both $M_2 <|\mu|$ and $M_2 >|\mu|$ cases.
The assigned values for the 
fundamental parameters in our numerical analysis indeed 
satisfy the assumption which went into the 
expressions (15)-(18).  
On the other hand, the switched behaviour of $M_{{\chi}^{0}_{3}}$ stems from the 
fact that  the  different
$\tan\beta$ contributions given by (17) compete against each other, and their roles are 
changed at a certain critical value. 
Namely,  there is a  critical value of 
the lighter chargino mass $M_{\chi^+_1}=130~\mbox{GeV}$
at which the $\tan\beta$-$M_{{\chi}^{0}_{3}}$ behaviour reverses.

Our analysis shows that  for the lower value of the lighter chargino mass  
($M_{\chi^+_1}=105~
\mbox{GeV}$), $M_2$  $(|\mu|)$ ranges in the $\sim 104-113~\mbox{GeV}$  
interval, as $|\mu|$ $(M_2)$ changes from $\sim 299-296~\mbox{GeV}$,  for $M_2 < |\mu|$ 
($M_2 > |\mu|$), as   $\tan\beta$
ranging from 5 to 50. On the other hand, for the higher value of the 
lighest chargino mass  ($M_{\chi^+_1}=160~\mbox{GeV}$), the corresponding values 
are: $M_2 \, (|\mu|)$: 164-175$~\mbox{GeV}$, and   
$|\mu| \, (M_2)$: 297-290  $\mbox{GeV}$.

In the CP violating case, the complementary  behaviour among  
the heavier neutralinos (i=3,4) can be observed  
in the sense that  while the CP violating points appearing  as local maxima 
for the ${\chi}^{0}_{3}$ mass, they turn out to be local minima for  ${\chi}^{0}_{4}$, 
for values of the lighter chargino mass in the  $M_2 < |\mu|$ regime.
This complementarity at the CP violating points for the heavier sector can easily
be seen from the expressions (17)-(18).
Such a  behaviour can also be observed for the lighter neutralinos
(i=1,2), but only
for  the lighter chargino case  in the  $M_2 > |\mu|$ regime.

\end{document}